\documentclass[twocolumn]{aastex631}
\usepackage{graphicx}
\usepackage{epstopdf}
\usepackage{natbib}
\usepackage{graphicx}
\usepackage{natbib}
\bibpunct{(}{)}{;}{a}{}{,}
\usepackage{booktabs}
\usepackage{array}

\newcommand{\ee}[1]{\mbox{${} \times 10^{#1}$}}
\newcommand{\eten}[1]{\mbox{$10^{#1}$}}

\newcommand{\degree}{\mbox{$^{\circ}$}}

\newcommand{\kms}{\mbox{km s$^{-1}$}}
\newcommand\cmv{\mbox{cm$^{-3}$}}

\newcommand{\x}{\mbox{${}\times{}$}}

\newcommand{\lsun}{\mbox{L$_\odot$}}
\newcommand{\msun}{\mbox{M$_\odot$}}

\newcommand{\tk}{\mbox{$T_K$}}

\newcommand{\vlsr}{\mbox{$v_{LSR}$}}
\newcommand{\n}{\mbox{$n$}}

\newcommand{\mean}[1]{\mbox{$\langle#1\rangle$}} 
\newcommand{\av}{\mbox{$A_V$}} 
\newcommand{\rinf}{\mbox{$r_{\rm{inf}}$}} 
\newcommand{\rout}{\mbox{$r_{out}$}} 

\newcommand{\hh}{\mbox{{\rm H}$_2$}}

\newcommand{\water}{H$_2$O}

\newcommand{\methanol}{\mbox{{\rm CH}$_3$OH}}

\newcommand{\hcop}{\mbox{HCO$^+$}}

\newcommand{\hccn}{\mbox{H$^{13}$CN}}

\newcommand{\jj}[2]{\mbox{$J = #1\rightarrow#2$}}

\newcommand{\go}{\mbox{$G_0$}}

\newcommand{\mstar}{\mbox{$M_{\star}$}}
\newcommand{\rstar}{\mbox{$R_{\star}$}}
\newcommand{\tstar}{\mbox{$T_{\star}$}}
\newcommand{\lstar}{\mbox{$L_{\star}$}}

\newcommand{\msunyr}{\mbox{M$_\odot$ yr$^{-1}$}}

\newcommand{\cseff}{\mbox{$c_{{\rm s,eff}}$}} 
\newcommand{\omegaz}{\mbox{$\Omega_0$}}

\newcommand{\tcol}{\mbox{$t_{\rm col}$}}
\newcommand{\router}{\mbox{$r_{\rm out}$}} 
\newcommand{\rinner}{\mbox{$r_{\rm in}$}} 

\newcommand{\thetacav}{\mbox{$\theta_{\rm cav}$}} 
\newcommand{\rhocav}{\mbox{$\rho_{\rm cav,0}$}} 
\newcommand{\rcav}{\mbox{$R_{\rm cav,0}$}} 
\newcommand{\thetaincl}{\mbox{$\theta_{\rm incl}$}} 
\newcommand{\mdisk}{\mbox{$M_{\rm disk}$}} 
\newcommand{\diskheight}{\mbox{$h_{\rm 100}$}} 
\newcommand{\hyperion}{\mbox{HYPERION}}
\newcommand{\spitzer}{\mbox{\rm{Spitzer}}}
\newcommand{\herschel}{\mbox{\rm{Herschel}}}
\newcommand{\alma}{\mbox{\rm{ALMA}}}
\newcommand{\wise}{\mbox{\rm{WISE}}}
\newcommand{\neowise}{\mbox{\rm{NEOWISE}}}
\newcommand{\jwst}{\mbox{\rm{JWST}}}

\shorttitle {Models of B335 }
\shortauthors{Evans et al.}

\begin{document}

\title{Models of Rotating Infall for the B335 Protostar}

\correspondingauthor{Neal J. Evans II}
\email{nje@astro.as.utexas.edu}

\author{Neal J. Evans II}
\affiliation{Department of Astronomy, The University of Texas at Austin,
2515 Speedway, Stop C1400, Austin, Texas 78712-1205, USA}
 
\author{Yao-Lun Yang}
\affiliation{RIKEN Cluster for Pioneering Research, Wako-shi, Satiama, 351-0198, Japan}

\author{Joel D. Green}
\affiliation{Space Telescope Science Institute, 3700 San Martin Dr., Baltimore, MD 02138, USA}

\author{Bo Zhao}
\affiliation{
Max-Planck-Institut f{\"u}r Extraterrestrische Physik (MPE), Giessenbachstr 1, D-85748 Garching, Germany}

\author{James Di Francesco}
\affiliation{National Research Council of Canada, Herzberg Astronomy and Astrophysics Reseach Centre, 5071 West Saanich Road, Victoria, BC V9E 2E7, Canada}

\author{Jeong-Eun Lee}
\affiliation{Department of Physics and Astronomy, Seoul National University, 1 Gwanak-ro, 4 Gwanak-gu, Seoul 08826, Korea}

\author{Jes K. J{\o}rgensen}
\affiliation{ Niels Bohr Institute, University of Copenhagen, {\O}ster Voldgade 5--7, 1350 Copenhagen K., Denmark}
\author{Minho Choi}
\affiliation{Korea Astronomy and Space Science Institute, 776 Daedeokdaero, Daejeon 305-348, Korea}
\author{Philip C. Myers}
\affiliation{Harvard-Smithsonian Center for Astrophysics, 60
  Garden Street, Cambridge, MA 02138, U.S.A.}
\author{Diego Mardones}
\affiliation{Departamento de Astronom{\'i}a, Universidad de Chile, Casilla 36-D, Santiago, Chile}

\begin{abstract}
Models of the protostellar source, B335, are developed using axisymmetric three-dimensional models to resolve conflicts found in one-dimensional models. The models are constrained by a large number of observations, including ALMA, Herschel, and Spitzer data.
Observations of the protostellar source B335 with ALMA show red-shifted absorption against a central continuum source indicative of infall in the HCO$^+$ and HCN $J = 4\rightarrow 3$ transitions.
The data are combined with a new estimate of the distance to provide strong constraints to three-dimensional \added{radiative transfer} models including a rotating, infalling envelope, outflow cavities, and a very small disk. The models favor ages since the initiation of collapse between 3\ee4 and 4\ee4 yr for both the continuum and the lines, resolving a conflict found in one-dimensional models. The models under-predict the continuum emission seen by ALMA, suggesting an additional component such as a pseudo-disk. The best-fitting model is used to convert variations in the 4.5 \micron\ flux in recent years into a model for a variation of a factor of 5-7 in luminosity over the last 8 years.
\end{abstract}


\section{Introduction}\label{intro}

Star formation occurs in a wide range of environments, from
isolated small clouds forming single stars to large dense clumps
forming massive stellar clusters in even larger molecular clouds.
The simplest of these systems
are isolated Bok globules
\citep{1947ApJ...105..255B},
where ``cloud," ``clump," and ``core" are synonymous. Among these,
B335 stands out for its isolation and simplicity.

\citet{1983ApJ...274L..43K} detected an infrared source in B335, the
first evidence for low-mass star formation in a Bok globule.
The first detection of infall from line profiles and detailed
comparisons to simple, inside-out collapse models were made
for B335
\citep{1993ApJ...404..232Z}
and subsequent works have continued to find general consistency
with such models
\citep{1995ApJ...448..742C, 2005ApJ...626..919E}.
These facts make B335 a natural target for fine-scale observations
with \alma.
\citet{2015ApJ...814...22E}
detected red-shifted absorption against the continuum in 
several lines with Cycle 1 ALMA observations, providing a
smoking-gun detection of infall.
\citet{2015ApJ...812..129Y}
showed that any disk in B335 has to be very small, suggesting
magnetic braking and/or a very young age, enhancing the interest in observations with better spatial resolution. In addition, B335 has a very compact region with
emission from complex organic molecules 
\citep{2016ApJ...830L..37I, 2019ApJ...873L..21I, 2022ApJ...935..136O}.

The distance to B335 has been
uncertain. For many years, the distance was taken to be 250 pc
\citep{1979PASJ...31..407T}.
\citet{2008ApJ...687..389S}
made the case that B335 was nearer, adopting a distance of 150 pc.
\citet{2009A&A...498..455O} estimated a distance to B335 using extinction
to nearby stars resulting in a distance between 90 pc and 120 pc. Most recent
work has assumed a distance of 100 pc. However, \citet{2009A&A...498..455O} noted that the southwestern rim of the globule is brightened in a deep U-band image. They considered whether this rim brightening could be caused by HD184982, an A2 star,
which then had a distance of 140 to 200 pc, based on Hipparcos. They instead concluded that the U-band excess was just the point-spread function from
the star, which was not in their field of view. 
Very recently, 
\citet{2020RNAAS...4...88W}
 has clearly established that 
B335 is indeed scattering light from
HD184982, which lies 242\arcsec\ west and 66\arcsec\ south of the B335
 submillimeter continuum source.
The GAIA DR3 \citep{2016A&A...595A...1G, 2021A&A...649A...1G}
 parallax to this star, HD184982, translates to a distance of
$164.5\pm 1.2$ pc.
More recently, optical absorption spectra
(S. Federman, personal communication)
obtained toward HD184982  indicate that the star
lies somewhat in front of the cloud, while another star, HD185176, at 247 pc distance,  clearly lies behind it. Conservatively, the distance to B335 must be between
164.5 and 247 pc, but the nebulosity connection to HD184982 argues that it
must be very close to that star.
We adopt a distance of 164.5 pc for this paper.
Consequently, the limit on the size of a Keplerian disk from
\citet{2015ApJ...812..129Y}
becomes $r < 16$ au.

Very recently, it has been realized that the source had varied substantially at 3.6 \micron\ and 4.5 \micron, as measured in the \wise\ 
\citep{2010AJ....140.1868W} W1 and W2 filters (C-H Kim et al. in prep.). The \neowise\ mission 
\citep{2014ApJ...792...30M}
has provided photometry of the source since MJD 56948 (2014, October 18) to add to the original \wise\ observation on MJD 55304 (2010, April 04).
The brightening began at least as early as MJD 56948 and was substantial (up to 2.5 mag), fading  by MJD 59000. 
The consequences for the luminosity history are explored in a later section. Most of the background information in the rest of this section is based on data that predate this variation, but those facts that may be affected will be noted.

Further evidence for variation can be found in studies of the outflow.
The source has a bipolar molecular outflow
\citep{1988ApJ...327L..69H}
oriented nearly E-W and in the plane of the sky, \added{ with an}
inclination angle with respect to the observer's line of sight of 87\degree
\citep{2008ApJ...687..389S}
with the blue lobe to the east of the source.
The highest resolution map of the inner part of the outflow
shows sharply defined cavity walls and a feature
interpreted as a molecular bullet, evidence for episodic ejection
\citep{2019A&A...631A..64B}. 
They estimated that the bullet was ejected about 1.7 yr before their 
2017 observations (MJD about 58051). 
The time delay calculation  assumed an inclination angle of
80\degree\ and a distance to B335 of 100 pc. At the current best distance of
164.5 pc, the outburst would have occurred 2.8 years earlier, around 
MJD 57027. The delay would be shorter if the inclination angle is 87\degree\
and the bullet was emitted along the outflow axis.
Analysis by C-H Kim (in prep.) suggests more recent ejections 
(between MJD 57570 and MJD 57844).
\citet{1992ApJ...395..494A} discovered a compact source at 3.6 cm, 
and suggested that it is a thermal jet.
Subsequent observations suggested variability on a timescale of
years 
\citep{2002AJ....124.1045R}
or perhaps even a month
\citep{2004RMxAA..40...31G}.

Variation on longer timescales is suggested by the presence of Herbig-Haro objects.
Herbig-Haro object HH119 has three main components, which lie east and west
of the infrared source
\citep{1992A&A...256..225R}. Two of these are 
about 9400 au away from the source,
scaled to the 164.5  pc distance.
Further HH objects were found by 
\citet{2007A&A...475..281G}.

The magnetic field in B335 has been studied  in the
near-infrared
\citep{2020ApJ...891...55K}, 
and with \alma\
\citep{2018MNRAS.477.2760M},
JCMT
\citep{2019ApJ...871..243Y},
and SOFIA
\citep{2021A&A...645A.125Z}
polarization observations.
\citet{2019ApJ...871..243Y}
found that the large scale magnetic field is aligned with the rotation
axis (E-W) to within about $10\degree$ on the plane of the sky, but
that the polarization vectors rotate by about 90\degree\ on small
scales, suggesting a magnetic pinch.
The \alma\ observations  
\citep{2018MNRAS.477.2760M}
probe the polarization on scales of 50 au to 500 au at 1.3 mm. Assuming
that the polarization results from aligned grains, they found an
equatorial field along the midplane (N-S) near the center but fields
along the outflow cavity walls, especially in the east (blue-shifted)
lobe. In addition, there was a peculiar patch of polarization in the 
north.
\citet{2018MNRAS.477.2760M}
found a good match to MHD models with relatively high
mass-to-flux ratios, producing a strong pinch in the mid-plane, but
cautioned that the strong polarization along the outflow in an edge-on
source like B335 could cause an overestimate of the pinch.
\citet{2020ApJ...893...54Y} argued for a more complex magnetic field
arrangement inside 160 au, perhaps as a result of a magnetic field
misaligned with the rotation axis. 
No evidence of ion-neutral drift was found on scales of 160 au
\citep{2018A&A...615A..58Y}.

\citet{2021A&A...653A.166C} have used rare CO isotopes to study the 
velocity field in the inner envelope. The complex pattern of line
profiles suggests the presence of accretion streamers along the cavity walls
as well as in the equatorial plane.
The kinematics very close to the forming star indicate a very small,
if any, Keplerian disk 
(\citealt{2015ApJ...812..129Y, 2019A&A...631A..64B, 2019ApJ...873L..21I,
2022ApJ...935..136O}).
\citet{2019ApJ...873L..21I} found a velocity gradient along a NW-SE direction
in a \methanol\ line, but they modeled it as part of an infalling rotating
envelope, rather than a Keplerian disk. 
\citet{2019A&A...631A..64B} found a N-S gradient in a different
\methanol\ transition of 0.9-1.4 \kms\ au$^{-1}$, but they also do not
find a good fit to a Keplerian model.

\citet{2021ApJ...918....2C} 
have used background stars to make a map of extinction that shows
a mostly spherical structure at low levels, but less extinction at
higher levels along the outflow lobes, leading to a north-south ``bow-tie"
appearance as extinction approaches the maximum measurable, $\av = 36$ mag.

\citet{2021A&A...650A.172J}
have detected compact D$_2$O emission toward B335; the abundance ratios of that
species to those of HDO and \water\ indicate that D$_2$O is inherited from the cold prestellar
core, rather than formed in situ.
\citet{2020ApJ...904...86C} have measured absorption features by ices toward
a star behind  B335, showing that water and CO ice are present.
The background star is 34\farcs60, or 5690 au in projected distance
from the central
source. The \spitzer\ spectrum of B335 itself
suggests deep ice features but the spectrum is highly uncertain, 
as discussed below.

Although observations of spectral lines on scales larger than
a few 1000 au have been consistently
well modeled by inside-out collapse models of originally isothermal
spheres 
\citep{1993ApJ...404..232Z,1995ApJ...448..742C, 2005ApJ...626..919E,
2015ApJ...814...22E},
with ages (defined as the time since collapse began in the inside-out
collapse model) of about $\tcol = 3-5\ee4$ yr, spherical models of
continuum observations consistently prefer a younger age or even
a power law all the way to the center
\citep{2002ApJ...575..337S,2011ApJ...728..143S}.
This tension between models suggests that re-analysis with 
new data and more flexible\deleted{, 3D} models is needed. The youngest
plausible age is the age of the CO outflow ($\tcol = 2\ee4$ yr),
based on 
\citet{1988ApJ...327L..69H}
scaled to a distance of 164.5 pc.
The newer, high-resolution maps typically do not cover the entire outflow, so
underestimate the age.

\section{Observations and Results}\label{obs}

Observations used to constrain the pre-outburst model include 
spectrophotometry with the
PACS and SPIRE instruments on \herschel. 
We also use reprocessed
archival \spitzer-IRS data, SCUBA data
\citep{2000ApJS..131..249S},
 and \alma\ observations from Cycle 1. 
Photometry at wavelengths  shorter and longer than the
\herschel\ data were obtained from, respectively,  
\citet{2008ApJ...687..389S}
and 
\citet{2000ApJS..131..249S}. 
Photometric points at 35 \micron\ and 70 \micron\ were obtained from
calibrated Spitzer and Herschel spectrophotometry. 
The infrared and submillimeter 
photometry is listed in Table \ref{phottab}.

ALMA data from a Cycle 3 project
that was continued into later cycles were obtained in
2016 and 2018, fortuitously providing data at different points in the 
luminosity evolution.
To constrain the variation in source luminosity, we also use data from WISE, NEOWISE, and SOFIA.

\subsection{Herschel Data}
The PACS observations were obtained by the DIGIT (Dust, Ice, and Gas in Time)
key program
\citep{2013ApJ...770..123G,2016AJ....151...75G}.
The SPIRE observations
were obtained by the COPS (CO in Protostars) open-time program 
\citep{2018ApJ...860..174Y}.
The instruments, observational methods, and data reduction have
been described in detail in those papers.
For B335, the spectra produced by the improved processing
\citep{2018ApJ...860..174Y}
agreed with photometry on images obtained
from the \herschel\ archive to within 35\% at 500 \micron,
and less than 24\% at shorter wavelengths
\citep{2018ApJ...860..174Y}.
The PACS spectrum was used to define a photometric point at 70 \micron,
while PACS photometry was used for longer wavelengths.

\subsection{Spitzer IRS data}\label{irsobs}

We used {\it Spitzer} IRS data from the archive
but did a new reduction. The field is dominated by scattered light
in the outflow lobes at short wavelengths
\citep{2008ApJ...687..389S}
making source extraction and sky subtraction difficult. Previous
reductions
\citep{2018ApJ...860..174Y}
showed artifacts and were not representative of the source.
The \spitzer\ IRS spectrum was produced using a mixed method that we describe here. We began with the basic calibrated data products from the \spitzer\ Heritage Archive. The Short-Low (SL2; 5-7.5 \micron, SL1; 7.5-14 \micron), Short-High (SH; 10-20 \micron) and Long-High (LH; 20-36 \micron) spectra were observed under AOR 3567360 using a raster mapping technique
\citep{2004ApJS..154..391W},
in which six positions were observed surrounding the source position - a pair of observations at the two nominal nod positions separated by 
$\sim1/3$ of a slit length, a second pair $+1/2$ the slit width in the perpendicular direction, and a third pair $-1/2$ the slit width in the perpendicular direction. This strategy was intended to guarantee that the source position would be observed even if the observation grid was not centered
on the source. In this case, the source was clearly detected in four of the six raster positions for SL2 (at $+6$ and $-4$ pixels from center for nod 1 and 2, respectively), SL1 (at $+6$ and $-4$ pixels from center), and LH (at $+1$ and $-1$ pixels from center). The source was not detected in any of the raster positions for SH, so we do not use those data.
The SL observations were taken with a slit oriented about 14\degree\
west of north, so perpendicular to the outflow.

We attempted to use local and more distant sky to subtract excess emission in SL1 and SL2, for which the slit is sufficiently long to identify sky, but found even the off-positions dominated by substantial diffuse emission that reduced the spectral signal-to-noise ratio. Furthermore, as we could not remove sky background from LH, we decided instead not to remove background.
The result will show the local shape of the SED well, but may need
to be scaled in absolute flux density.

We describe the individual extraction methods for each module, using the SMART package
\citep{2010PASP..122..231L}. 
For SL2 and SL1, we extracted 6 pixel wide (10\farcs8) 
``fixed-column'' apertures centered on the source peak in each nod in 
each of the four positions in which the source was detected. 
We did not perform any sky subtraction. We averaged the four spectra by
module, removed the order edges, and produced a single combined SL2$+$SL1 spectrum.
The SL1 aperture was 3\farcs7 by 10\farcs8 (6 pixels at 1\farcs8 per pixel) and SL2 was 3\farcs6 by 10\farcs8, while LH was 11\farcs1 by 22\farcs3.
For LH, we extracted the full aperture (22\farcs5  wide), and clearly detected the source peak centered at a position consistent with SL1/SL2 in four of the six raster positions. We averaged those four position spectra, removed the order edges, and produced a single combined LH spectrum.
This spectrum  was scaled down by a factor of 2 to match the photometric point
at 24 \micron.
We used the scaled LH data 
to create a pseudo-photometric point at 35 \micron,
where no photometry was available, and we used the SL1 data
to create such a point at 9.7 \micron. These are definitely
uncertain, but provide useful constraints.

The J2000 positions of the peak  were
19:37:00.7, +7:34:11 for SL2 ($\lambda \sim 7 \micron$) and
19:37:01.0, +7:34:08 for LH ($\lambda \sim 29 \micron$).
Both are consistent with the \alma\ peak (\S \ref{almaobs})
within the 3\arcsec\
uncertainties in the \spitzer\ peak position.

\subsection{The Spectral Energy Distribution and Radial Profiles}

\begin{figure}
\center
\includegraphics[scale=0.3, angle=0]{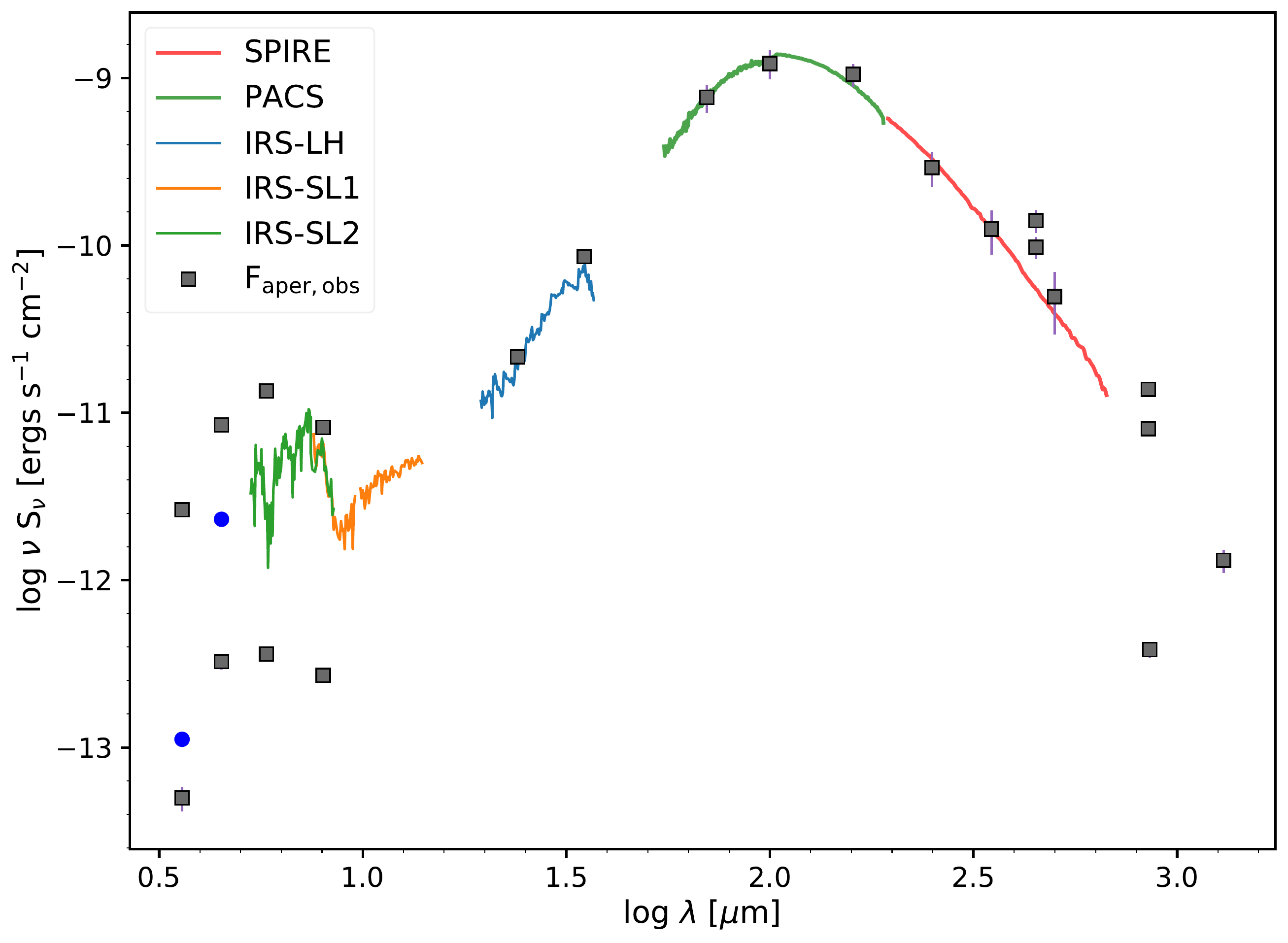}
\caption{
The observed SED using only data from MJDs before 57000 (Table \ref{phottab}).
The solid points are photometric data, while the continuous lines are
spectrophotometry from  Herschel-PACS (green), 
and Herschel-SPIRE (red).The Spitzer IRS data are shown in green (SL2),
orange (SL1), and blue (LH). 
\added{The blue circles at 3.6 and 4.5 \micron\ are the WISE data from Table
\ref{phottab}.}
The black boxes with error bars are photometric
or pseudo-photometric points based on the IRS and PACS spectra to 
fill in the gap between 24 \micron\ and 100 \micron. Different photometric points
at the same wavelength reflect different apertures. The lowest point around 850 \micron\ is the Cycle 1 ALMA flux in a 2\arcsec\ aperture.
}
\label{seddata}
\end{figure}

\begin{deluxetable}{cccccl}
\tabletypesize{\footnotesize}
\tablewidth{0pt}\tablecaption{Photometry for B335}
\tablehead{
\colhead{$\lambda$} &
\colhead{$S_\nu$} & 
\colhead{$\sigma(S_\nu)$} &
\colhead{Aperture} &
\colhead{MJD} &
\colhead{Notes} \\
\colhead{(\micron)} &
\colhead{(Jy)} &
\colhead{ (Jy) } &
\colhead{(arcsec) } &
\colhead{  } &
\colhead{  }  \\
\tableline
\colhead{(1)} &
\colhead{(2)} &
\colhead{(3)} &
\colhead{(4)} &
\colhead{(5)} &
\colhead{(6)}
}
\startdata
3.6         &  6.00\ee{-5} & 1.0\ee{-5}       &  2.4   & 53116 &  1     \\
3.6         &  1.34\ee{-4} & 1.3\ee{-5}       &  5.84  & 55304 &  2     \\  
3.6         &  3.16\ee{-3} & 3.0\ee{-4}       &  24    & 53116  &  1     \\ 
4.5         &  4.90\ee{-4} & 5.0\ee{-5}       &  2.4   & 53116   &  1     \\
4.5         &  3.46\ee{-3} & 3.5\ee{-4}       &  6.47  & 55304   &  2     \\
4.5         &  1.27\ee{-2} & 1.0\ee{-3}       &  24    & 53116   &  1     \\
5.8         &  7.00\ee{-4} & 7.0\ee{-5}       &  2.4   & 53116   &  1     \\
5.8         &  2.61\ee{-2} & 2.5\ee{-3}       &  24    & 53116   &  1     \\
8.0         &  7.20\ee{-4} & 7.0\ee{-5}       &  2.4   & 53116   &  1     \\
8.0         &  2.18\ee{-2} & 2.0\ee{-3}       &  24    & 53116  &  1     \\
24.0        &  1.73\ee{-1} & 1.6\ee{-2}       &  26    & 53293   &  1     \\
35.0        &  1.00\ee{0}  & 1.0\ee{-1}       &  22.5  & 53292   &  3     \\
70.0        &  1.79\ee{1}  & 6.0\ee{0}        &  24.8  & 55513   &  4     \\
100.0       &  4.07\ee{1}  & 8.1\ee{0}        &  24.8  & 55513   &  5     \\
160.0       &  5.61\ee{1}  & 8.7\ee{0}        &  24.8  & 53116   &  5    \\
214.0       &  3.69\ee{1}  & 2.0\ee{-1}       &  18.2  & 58044   & 6   \\
250.0       &  2.43\ee{1}  & 5.7\ee{0}        &  22.8  & 55513   &  5     \\
350.0       &  1.46\ee{1}  & 4.3\ee{0}        &  28.6  & 55513   &  5     \\
450.0       &  1.46\ee{1}  & 2.2\ee{0}        &  40    & 50880   &  7     \\
450.0       &  2.11\ee{1}  & 3.3\ee{0}        &  120   & 50880   &  7     \\
500.0       &  8.24\ee{0}  & 3.4\ee{0}        &  39.1  & 55513   &  5     \\
850.0       &  2.28\ee{0}  & 1.2\ee{-1}       &  40    & 50880   &  7     \\
850.0       &  3.91\ee{0}  & 2.2\ee{-1}       &  120   & 50880   &  7     \\
857.0       &  1.1\ee{-1}  & 1.2\ee{-2}       &  2.0   & 56774   &  8     \\
1300.0      &  5.70\ee{-1} & 9.0\ee{-2}       &  40    & 50880   &  7     \\
\enddata
\tablecomments{ 1. Spitzer photometry 
\citep{2008ApJ...687..389S}; 
2. WISE data;
3. Based on calibrated 
{\it Spitzer} spectrum;
4. Based on calibrated Herschel-PACS spectrum; 
5. Herschel-PACS or Herschel-SPIRE
\citep{2016AJ....151...75G};
6. SOFIA-HAWC$+$ 
 (Zielinaky et al. 2021);
7.  JCMT-SCUBA
\citep{2000ApJS..131..249S};
\added{8. ALMA Cycle 1 from Table \ref{tabalmares}.}
}
\label{phottab}
\end{deluxetable}

The resulting SED is shown in Figure \ref{seddata}, where we plot only
data from before MJD 57000.
Almost all these data were taken before the luminosity increase, so
we refer to this set as the quiescent SED.
The continuous lines are spectrophotometric data from 
\spitzer\ IRS, with the current reduction (\S \ref{irsobs}),
along with the data from \herschel-PACS, and \herschel-SPIRE 
\citep{2018ApJ...860..174Y}.
The black points with error bars are photometric data, listed in Table \ref{phottab}.
The integration over the SED yields an {\it observed} luminosity
of 1.36 \lsun, higher than estimated by
\citet{2015ApJ...814...22E},
due to the new distance and details of data chosen to constrain the integration.
We favored the calibrated spectrophotometry over the photometric data where
the wavelengths overlapped because of the better sampling of wavelengths.
In general, the two agree well. If we use the higher values of photometry
where they disagree, the observed luminosity  increases to 1.76 \lsun, but
we consider that as a strong upper limit.
Because of the edge-on geometry, the actual luminosity needed to
match the observations is higher (\S \ref{3dmodels}), and the
final determination of the actual luminosity depends only on
matching the modeled and observed photometric points.

\begin{figure}
\center
\includegraphics[scale=0.3, angle=0]{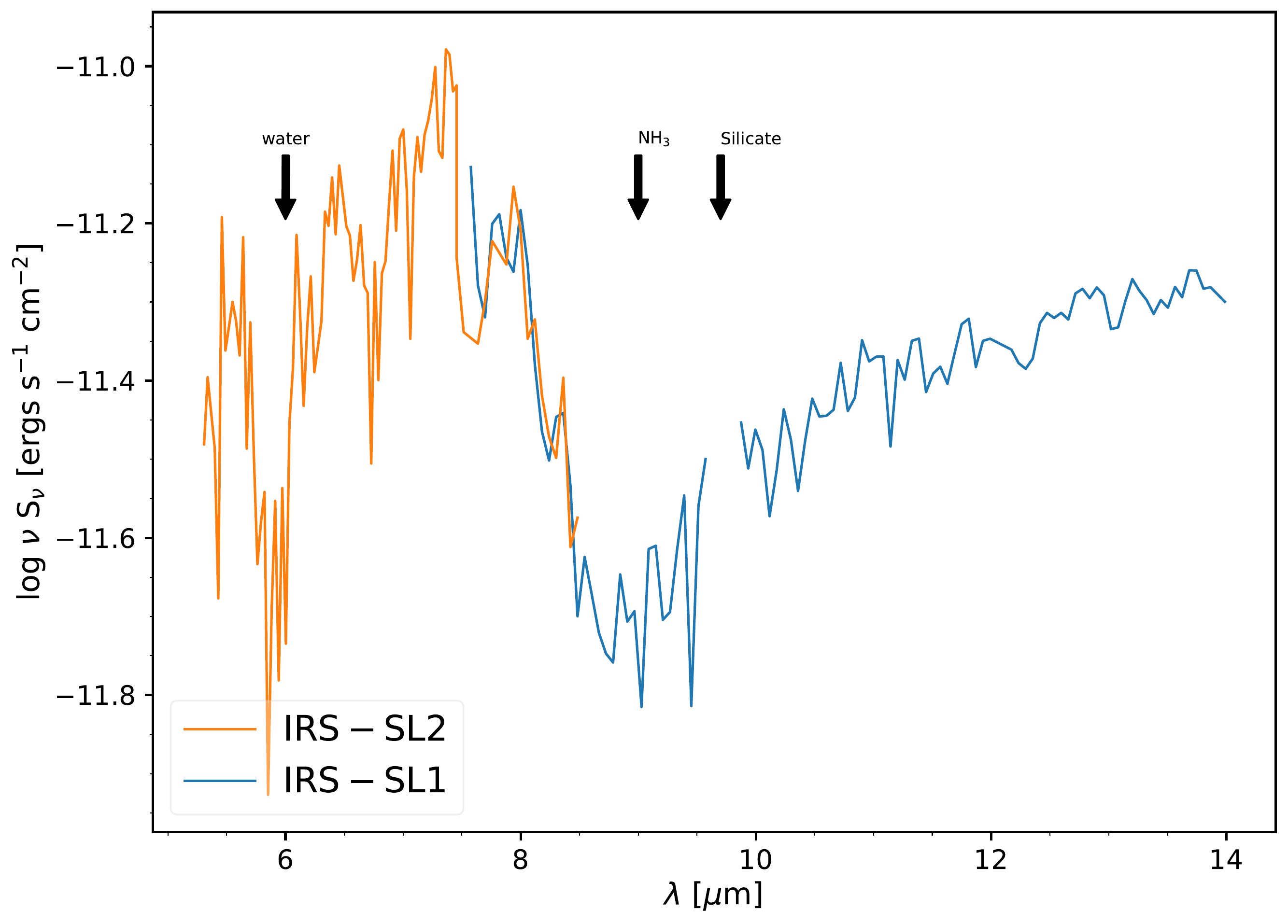}
\caption{
The spectrum from 5 \micron\ to 14 \micron\ from {\it Spitzer} IRS after
re-reduction. The wavelengths of key spectral features are
indicated by arrows.}
\label{spitzerobs}
\end{figure}

Figure \ref{spitzerobs} zooms in on the spectral region from 5 \micron\ to
14 \micron. The spectrum is noisy but quite remarkable. The silicate
feature is quite shallow for such an embedded source, while the
ice features at 6.0 \micron\ and 6.8 \micron\ are very deep. The
deepest absorption is around 9.0 \micron, a significantly shorter
wavelength than the usual maximum absorption by silicate grains. 
Finally, there seems to be broad absorption over the 13-14 \micron\
region of the water libration mode. This spectrum, while uncertain, is
different from all the spectra in 
\citet{2008ApJ...678..985B}.
The fact that we could not do sky subtraction may affect the spectrum.
In particular, the shallow silicate feature could arise in part from
scattered silicate emission.
Improved spectra should be available from approved observations with 
\jwst.

In addition to the SED, we use the radial profiles of emission
at 450 \micron\ and 850 \micron\ 
\citep{2000ApJS..131..249S},
which are sensitive to the radial variation in the density and
hence the evolutionary stage. These profiles provide the strongest constraint
on the age of the system.

\subsection{\alma\ data}\label{almaobs}

\alma\ data were obtained as part of Cycle 1 project 2012.1.00346.S
(PI N. Evans) on 27 April 2014 (UT) in the C32-3 configuration
of \alma\ Early Science. The details of the observations and
data reduction have been described by 
\citet{2015ApJ...814...22E}.
In that paper, the primary spectral line
targets (HCN, \hccn, and \hcop\ \jj43\ lines and CS \jj76\ line)
were presented and the HCN and \hcop\ lines were used to detect
infalling motions. 
The clean beam for Cycle 1 was $490 \times 440$ mas at $PA = -76\fdg9$.

\begin{deluxetable*}{l l r r r r l}
\tabletypesize{\footnotesize}
\tablewidth{0pt}\tablecaption{ALMA Observations}
\tablehead{
\colhead{ID} & 
\colhead{Date} & 
\colhead{Antennas} & 
\colhead{B(min)} & 
\colhead{B(max)} & 
\colhead{PWV} & 
\colhead{Notes} \\
\colhead{ } &
\colhead{ } &
\colhead{ } &
\colhead{(m)} &
\colhead{(m)} &
\colhead{(mm)} &
\colhead{ } 
}
\startdata
uid\_A002\_Xb57bb5\_X451  & 2016-07-17 & 41  & 15  & 1090 &  0.57 &   \\  
uid\_A002\_Xb5aa7c\_X4c04 & 2016-07-22 & 38  & 15  & 1090 &  0.52 &   \\ 
uid\_A002\_Xb5bd46\_X1b74 & 2016-07-23 & 37  & 16  & 1110 &  0.67 & 1 \\
uid\_A002\_Xb5bd46\_X2404 & 2016-07-23 & 37  & 16  & 1110 &  0.66 & 2 \\
uid\_A002\_Xd1daeb\_X2e93 & 2018-09-11 & 44  & 15  & 1213 &  0.73 &   \\ 
uid\_A002\_Xd21a3a\_X1d4a & 2018-09-18 & 47  & 15  & 1397 &  0.67 & 3 \cr
\enddata
\tablecomments{
 1. $>50$\% of antennas had phase fluctuations  over limit. \\
2. 18 antennas exceeded limit on phase fluctuations. \\
 3. 18 antennas had high phase rms, but were usable.
}
\label{tabalmaobs}
\end{deluxetable*}

\alma\ data with higher spatial resolution were obtained with the same
frequency set-ups in Cycle 3 (Table \ref{tabspw})
as part of project 2015.1.00169.S and Cycle 4
 2016.1.00069.S (PI. N. Evans). 
Six execution blocks were obtained, four in 2016, which did not meet the
noise specifications, and two in 2018; when combined, the data met the noise 
specifications. The properties of the observations are listed in Table
\ref{tabalmaobs}, including number of antennas, minimum and maximum baselines, and precipitable water vapor (PWV), along with notes on issues.
The properties of the lines observed are given in Table \ref{tabspw}.
The spectral resolution was 122.07 kHz, except for \hccn, for which it
was 488.28 kHz. The equivalent values of velocity resolution are listed
in Table \ref{tabspw}.

The data were calibrated and imaged using CASA 
\citep{2007ASPC..376..127M}
version 5.4.0. 
Observations of strong lines were subject to a calibration error when
all these data were obtained that could artificially weaken the lines.
The effect is roughly equal to the single-dish line temperature divided by the
system temperature.
The single dish lines
\citep{2005ApJ...626..919E}
have temperatures less than 3 K, versus typical system temperatures during
the \alma\ observations of about 180 K, so the effect of calibration error
should be less than 1.7\%, negligible given other uncertainties.


\begin{deluxetable*}{l r r r r c }
\tabletypesize{\footnotesize}
\tablewidth{0pt}\tablecaption{ALMA Spectral Windows}
\tablehead{
\colhead{Transition} &
\colhead{Frequency} &
\colhead{$\delta v$} &
\colhead{$E_{u}$} &
\colhead{$A_{ul}$} &
\colhead{Sensitivity} \\
\colhead{} &
\colhead{(GHz) } &
\colhead{(\kms) } &
\colhead{(K)} &
\colhead{(s$^{-1}$)} &
\colhead{(K/(Jy/bm)$^{-1}$)} \\
\tableline
\colhead{(1)} &
\colhead{(2)} &
\colhead{(3)} &
\colhead{(4)} &
\colhead{(5)} &
\colhead{(6)}
}
\startdata
CS \jj76\      & 342.88285030  & 0.107 & 65.8    & 8.4\ee{-4}  & 191 \\
\hccn\ \jj43\  & 345.33976930  & 0.424 & 41.4    & 1.90\ee{-3} & 214 \\
HCN \jj43\     & 354.50547590  & 0.103 & 42.5    & 2.05\ee{-3} & 202 \\
\hcop\ \jj43\  & 356.73422300  & 0.103 & 42.8    & 3.57\ee{-3} & 200  \\
\enddata
\tablecomments{
K was determined from the clean beam and frequency for each line. 
}
\label{tabspw}
\end{deluxetable*}

After learning of the  outburst, we separated the 2016 data from the 2018 data and reduced each separately and refer to them as Cycle3-16 and Cycle3-18. The delay between observations was extremely fortuitous as the lines in 2018 were much stronger, consistent with the brightening seen in the \wise\ data. The description of the data reduction below applies to both sets of data, except when noted.

Images were made with 0\farcs025 cells. Briggs weighting with a robust parameter of 0.5 was used and the task tclean used auto-multithresh for the clean mask, 
typically with 2 to 3 major cycles.
As explained in 
\citet{2015ApJ...814...22E},
removal of the continuum in the uv-plane can cause distortions in spectra 
with both emission and absorption. The lines and continuum were separated
after cleaning with the task imcontsub and spectra were extracted with the
task specflux with the region defined to be the size and orientation of the
continuum source before deconvolution. The continuum was subtracted in the image plane after careful specification of the line-free channels.
A separate reduction with the continuum removed in the uv-plane
produced nearly identical results, but with a poorer baseline
removal. Self-calibration was used on the continuum and applied to the lines.
The clean beam for the 2016 data after self-calibration was $200 \times 150$ mas at
$PA = -74\fdg1$.
The clean beam for the 2018 data after self-calibration was $240 \times 190$ mas at
$PA = -63\fdg1$.
The sensitivity, denoted $K$ in kelvin-(Jy/bm)$^{-1}$
and computed from the beam and frequency for each line, is given in Table
\ref{tabspw} for the 2018 data.

\begin{figure}
\center
\includegraphics[scale=0.35, angle=0]{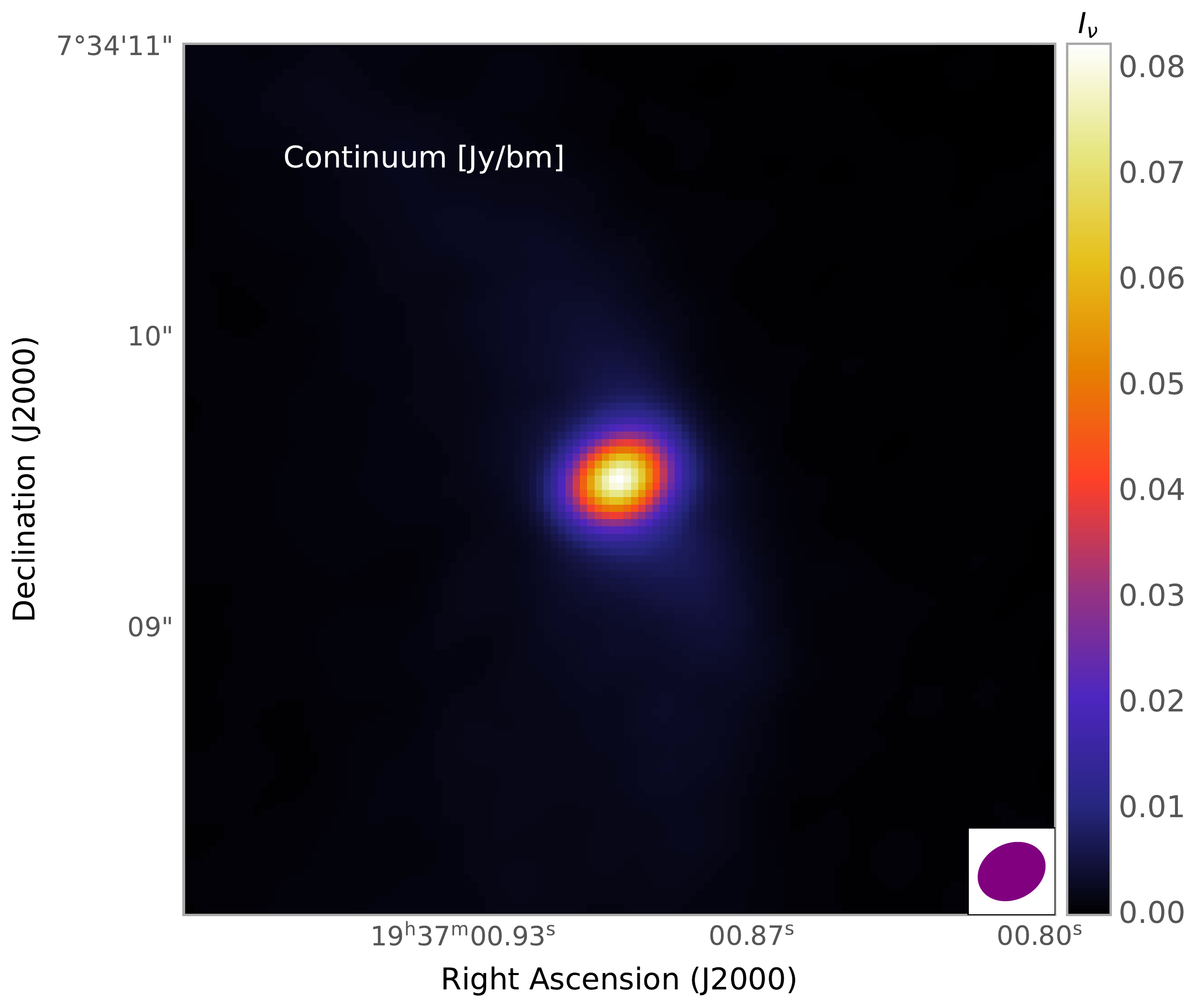}
\caption{
The continuum image 
from the ALMA Cycle 3 2018 data. The beam is indicated in the lower right corner.
}
\label{almacont}
\end{figure}

The continuum image from the Cycle 1 observations was presented in
\citet{2015ApJ...814...22E},
showing a compact central source and faint arcs from the outflow cavity.
The Cycle 3 data with better spatial resolution showed many lines of
complex organic species, which are presented in a separate paper.
They and the main target lines were avoided in constructing the
continuum image in Figure \ref{almacont}.

\begin{deluxetable*}{l l c c c c c c c}
\tabletypesize{\footnotesize}
\tablewidth{0pt}\tablecaption{ALMA Source Properties}
\tablehead{ 
\colhead{Dataset} &
\colhead{MJD} &
\colhead{$\theta_{max}$} &
\colhead{$\theta_{min}$} &
\colhead{PA} &
\colhead{$I_\nu$ } &
\colhead{$S_\nu$ (fit) }  &
\colhead{$S_\nu$ (2\arcsec )} &
\colhead{Species} \cr
\colhead{ } &
\colhead{ } &
\colhead{(mas)} &
\colhead{(mas)} &
\colhead{(degree)} &
\colhead{(mJy/bm) } &
\colhead{(mJy)} &
\colhead{(mJy)} &
\colhead{ } 
}
\startdata
Cycle1    & 56774 & $491\pm 49$    &  $258\pm 59$  & $9.6\pm 10.1$  & $55.1\pm 2.7$  & $93.5\pm 6.7$  & $110\pm 12$ & Cont. \\
Cycle3-16 & 57586 & $100.1\pm 7.2$ & $97.3\pm 8.3$ & $43\pm 79$     & $89.6\pm 1.0$  & $118.5\pm 2.0$ & $160\pm 8$ & Cont. \\
Cycle3-18 & 58379 & $186.0\pm 19$  &  $179\pm 22$  & $175\pm 79$    & $75.3\pm 2.0$  & $129.8\pm 5.1$ & $217.5\pm 8.6$ & Cont. \\
Cycle3-18 & 58379 & $789\pm 54$    &  $467\pm 34$  & $72.1\pm 5.3$  & $2.75\pm 0.16$ & $24.8\pm 1.6$  & $28.82\pm 0.63$ & \hcop\ \\
Cycle3-18 & 58379 & $259.3\pm 9.6$ & $228.3\pm 34$ & $103\pm 16$    & $6.45\pm 0.11$ & $14.51\pm 0.33$ & $16.83\pm 0.86$ & HCN \\
Cycle3-18 & 58379 & $188.6\pm10.7$ & $156.0\pm8.8$ & $122\pm 17$    & $6.04\pm 0.11$ & $9.64\pm 0.27$  & $8.04\pm 0.72$ & \hccn\ \\
Cycle3-18 & 58379 & $305\pm 26$    & $238\pm 19$   & $119\pm 19$    & $1.658\pm 0.072$ & $3.89\pm 0.23$  & $5.31\pm 0.24$ & CS \\
\enddata
\tablecomments{For the lines, the units are mJy-\kms or mJy/bm-\kms.}
\label{tabalmares}
\end{deluxetable*}

The J2000 peak position for continuum emission, averaged between 2016 and 2018, is
19:37:00.8967, $+$07:34:09.517 with a difference of only 50 mas,
consistent within uncertainties with the Cycle 1 peak position.
The deconvolved FWHM major and minor axes, peak intensity, and total flux 
based on an elliptical Gaussian fit are given in Table \ref{tabalmares}.
In addition, we tabulate the total flux density within
a 2\arcsec\ circular area, representing the flux density arising from a region
of diameter 329 au. 

\begin{figure*}
\center
\includegraphics[scale=0.4, angle=0]{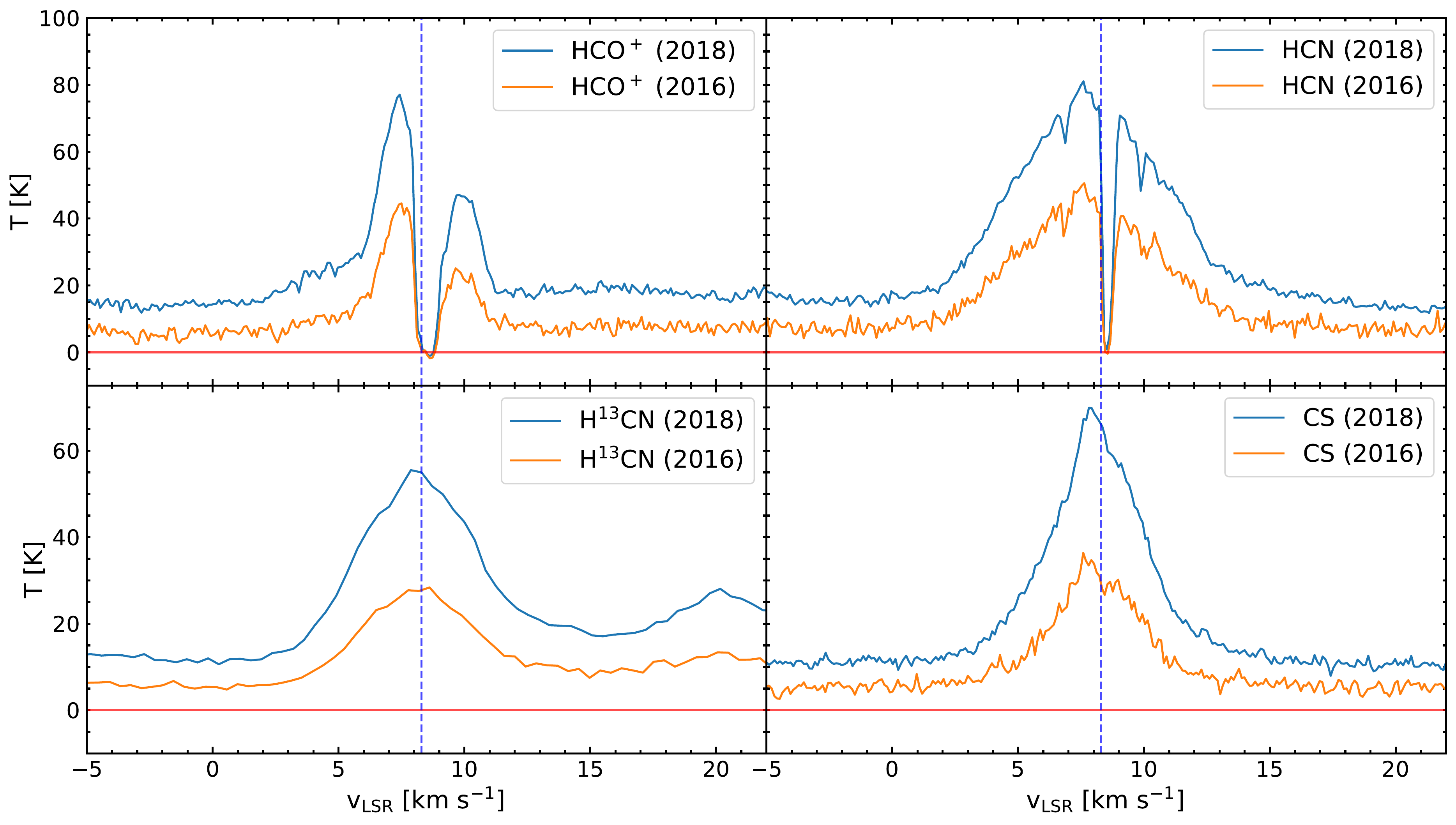}
\caption{
Spectra of the \hcop, HCN, CS, and H$^{13}$CN lines
in an 0\farcs2 diameter circle centered on the continuum source
for both the 2016 and 2018 data
before the continuum was removed in the image plane. 
The source velocity is shown as a vertical dashed line.
}
\label{4spec}
\end{figure*}

The  spectra of the four lines (Figure \ref{4spec}) from the 2016 and 2018 \alma\ data, in an 0\farcs2 diameter aperture  before removal of the continuum in the image plane, show
that the absorption of the continuum by the \hcop\ and HCN lines is essentially complete and clearly shifted to the red of the source
velocity of 8.3  \kms\
\citep{2005ApJ...626..919E}.
The HCN transition has hyperfine structure
\citep{2002ZNatA..57..669A, 2016MNRAS.459.2882M}, but 96\%
of the line strength lies in three hyperfine components within 0.14 \kms\
of the center frequency that we use to assign velocities. There are two very weak
components (2\% each) at $+1.34$ \kms\ and $-1.67$ \kms.
\hccn\ also has hyperfine structure
\citep{2004ZNatA..59..861F};
again, 96\% of the line strength lies within $0.11$ \kms of our center
frequency, with very weak (2\%) features at $+1.39$ \kms\ and $-1.72$ \kms.
 The peak at velocities below the blue dip in the HCN spectrum is
at 6.6 \kms, close to the offset expected for the very weak hyperfine
component, but the corresponding velocity for the other weak hyperfine
component (9.6 \kms) corresponds only to the shoulder to the blue of the
red dip, not to the secondary peak. No dips or secondary peaks are visible
in the \hccn\ spectrum, but some shoulders could plausibly be related to 
the weak hyperfine components.

The CS line shows a local dip near 8.3 \kms, but not absorption below the continuum
at the source velocity, while the \hccn\ peaks at the source velocity.
The rise at high velocities in the \hccn\ spectrum is caused by lines of
COMs, which also partly contribute to the \hccn\ line profile, along with a
possible contribution from a line of SO$_2$ at 356755.19 MHz. The 
smoother line profile of the \hccn\ observations results from the lower
spectral resolution chosen for that band.  There are additional dips in the HCN
spectrum at $6.9$ \kms\ and $9.9$ \kms.

For comparison to models, the continuum was removed, choosing line free channels to fit the continuum. First order
baselines were used, except for \hccn\ and HCN, for which 
there were too few line-free regions, so a zeroth order baseline was used.
We also removed the continuum in the uv-plane for images of
the extended emission, and these images are used for some purposes.

\begin{figure*}
\center
\includegraphics[scale=0.30, angle=0]{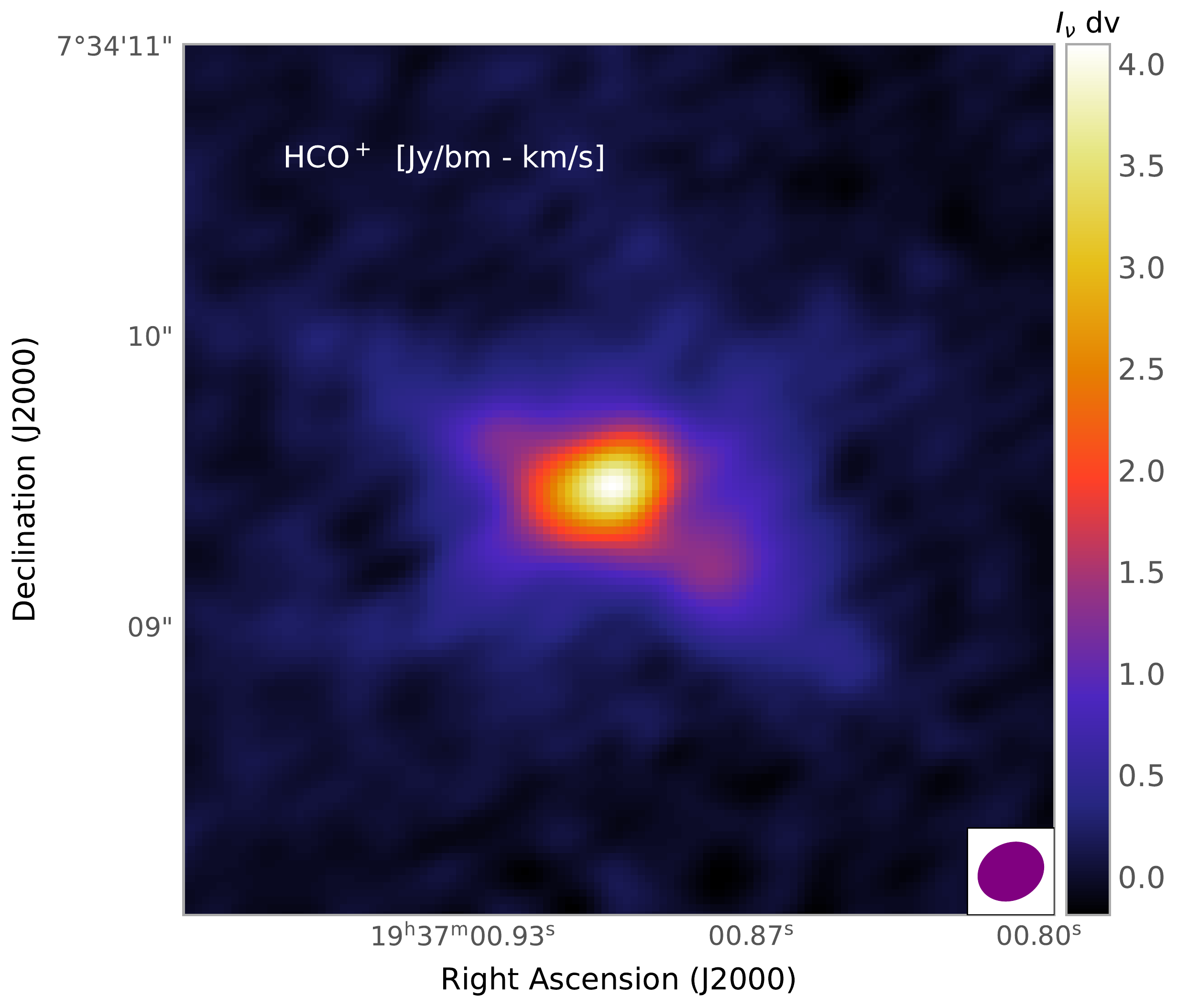}
\includegraphics[scale=0.30, angle=0]{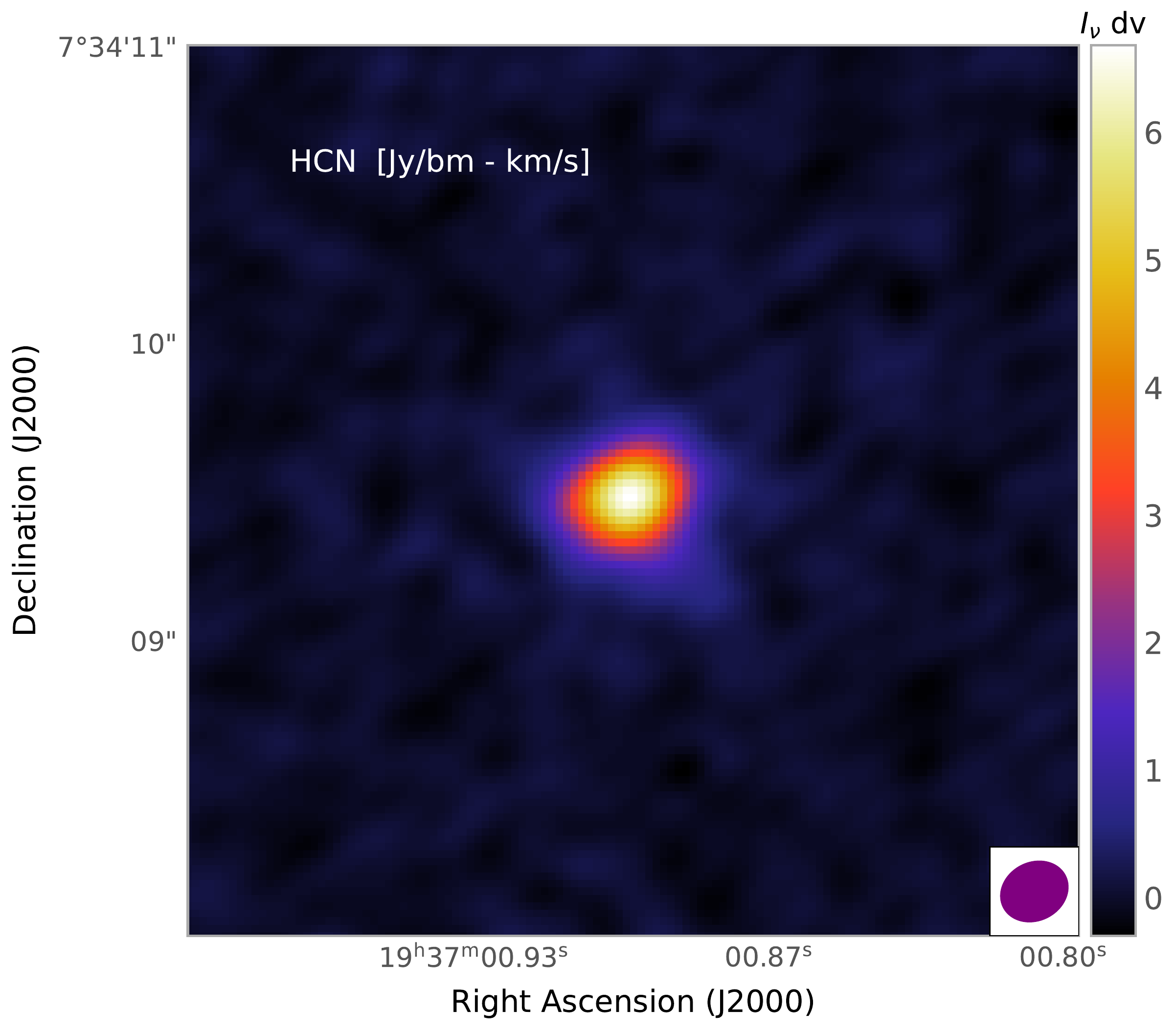}
\includegraphics[scale=0.30, angle=0]{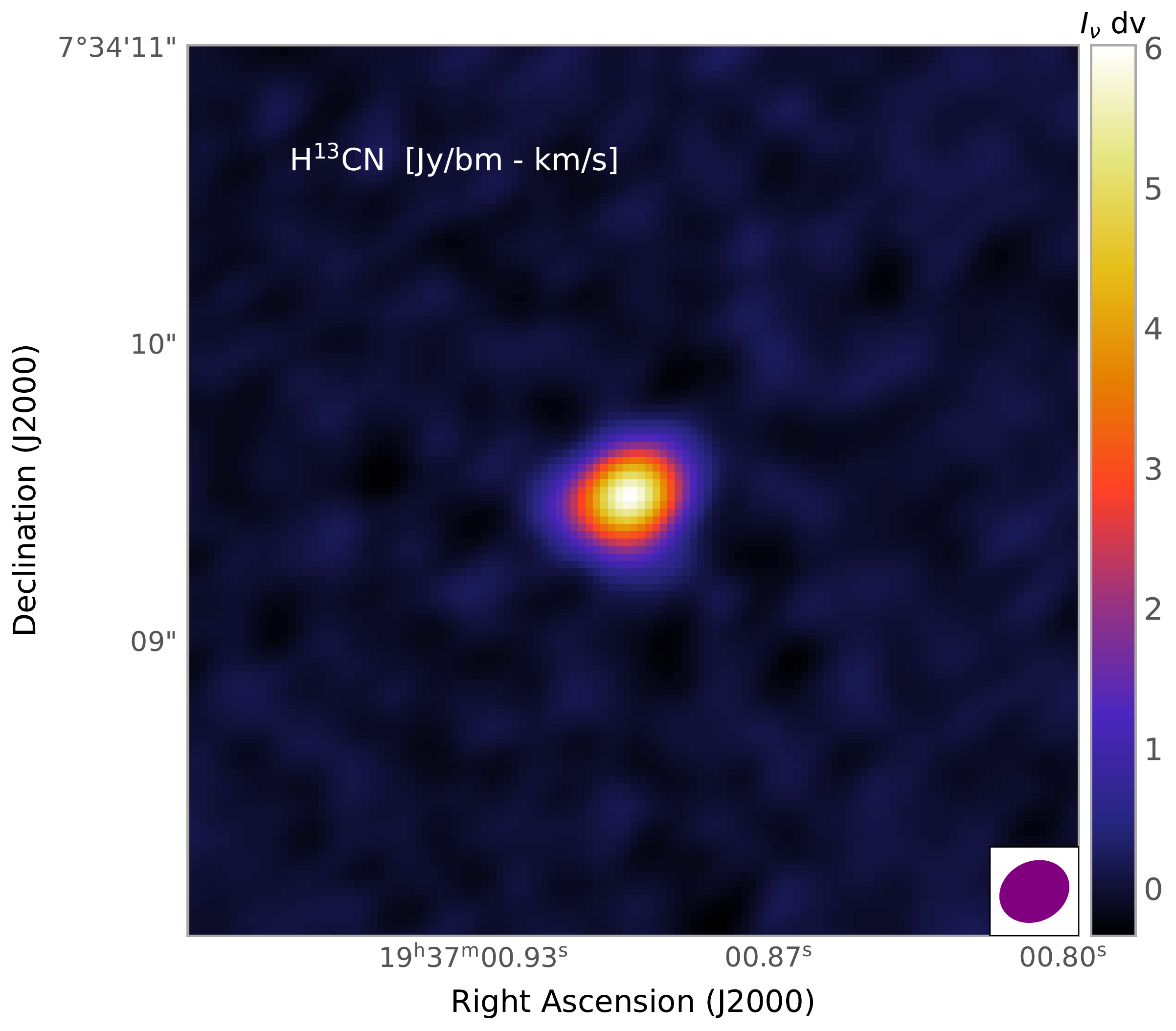}
\includegraphics[scale=0.30, angle=0]{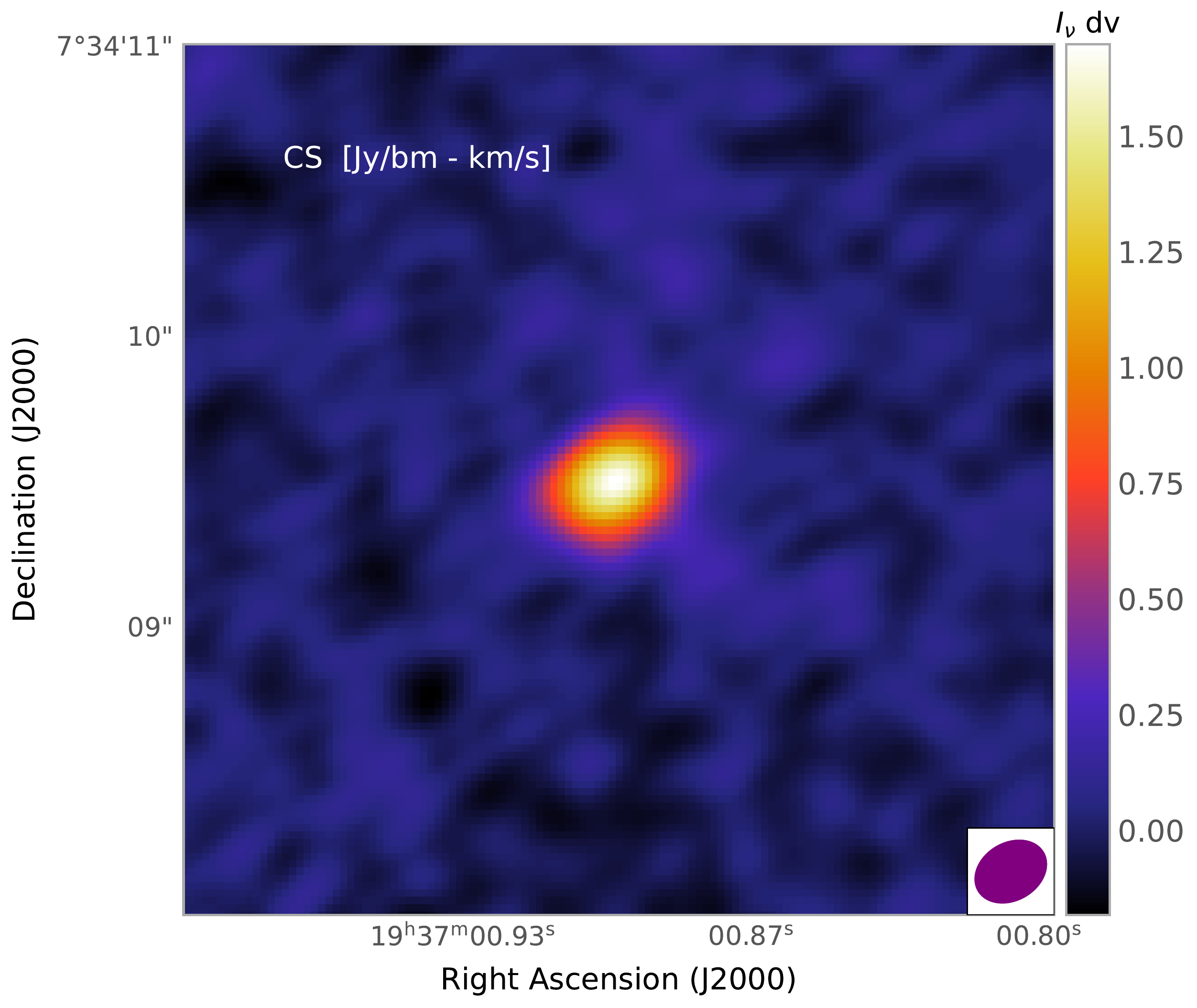}
\caption{
The \hcop, HCN, \hccn, and CS integrated intensity images  
from the ALMA Cycle 3 2018 data. The velocity intervals for integration were 1-14 \kms\ for \hcop, $-3$ \kms\ to 21 \kms\ for HCN, and 1-12 \kms\ for \hccn, and 1 to 17 \kms\ for CS. The beam is indicated in the lower right corner.
}
\label{almalineimages}
\end{figure*}

The intensity, integrated over the line, is shown in Figure \ref{almalineimages}.
The presence of other lines required tuning of the velocity interval, as indicated in the caption. The integrated intensity maps were fitted to 
Gaussians, and the fitted peak positions were the same for all lines to within
0\farcs04, or 20\% of the beam size. The average peak position of the lines 
agreed with that of the continuum to 0\farcs025, about 10\% of the beam size.
The fits were all compact, but they had different sizes, as indicated in  
Table \ref{tabalmares}. Except for \hcop, they are compact ellipses. The integrated intensity map of \hcop\ shows extensions likely to be the edges of the outflow cavities in addition to the compact ellipse.

\subsection{Outflow Structures}

\begin{figure}
\center
\includegraphics[scale=0.35, angle=0]{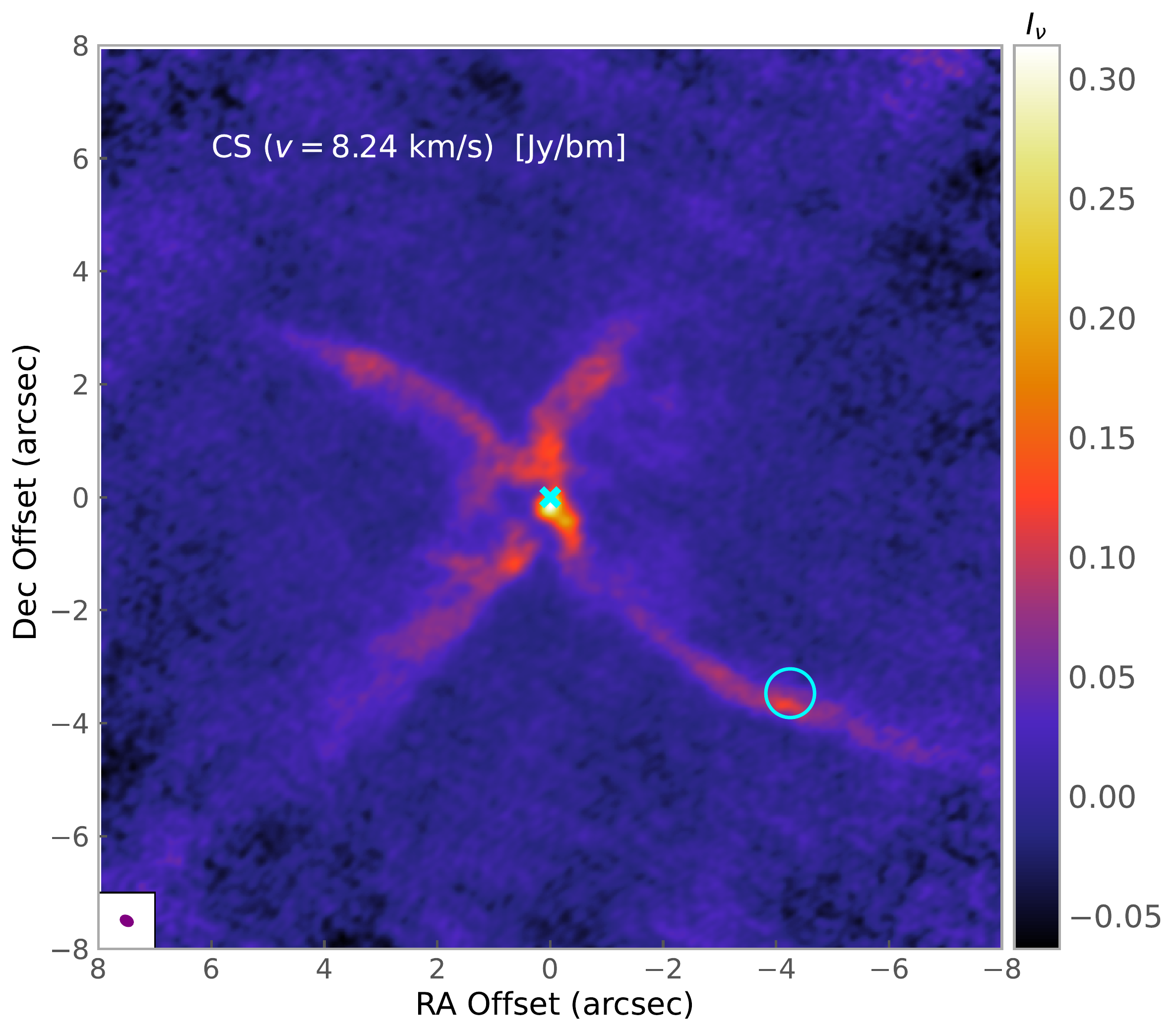}
\caption{
The CS image at $v = 8.24$ \kms\ from 2018. The cavity walls are most apparent at this velocity and this molecular transition. The image is taken from the data with the continuum removed in the image plane. \added{The X marks the continuum peak position, while the cyan circle indicates the location of the ``red blob."} The beam is indicated in the lower left corner.
}
\label{CSwalls}
\end{figure}

In addition to the spectra toward the continuum source, the \alma\ Cycle 3
data reveal some interesting features about the outflow cavity. Partial
arcs can be seen in several lines/velocities, but the outflow is prominent in neither the integrated intensity images nor in channels centered on outflow velocities. The most complete image
of the cavity walls is that of the CS \jj76\ transition at $v = 8.24$ \kms,
as shown in Figure \ref{CSwalls}. We interpret these arcs as the limb-brightened
top and bottom of the cavity walls because the velocity is close to the rest 
velocity and both blue and red lobes are clearly visible. The structure in
Figure \ref{CSwalls} is similar to the polarized continuum intensity 
at 0.87 mm in
figure 1 of \citet{2020ApJ...893...54Y} but shows the west lobe more
clearly. This structure is visible in all three epochs, but it is most prominent
in the 2018 data.

\begin{figure}
\center
\includegraphics[scale=0.23, angle=0]{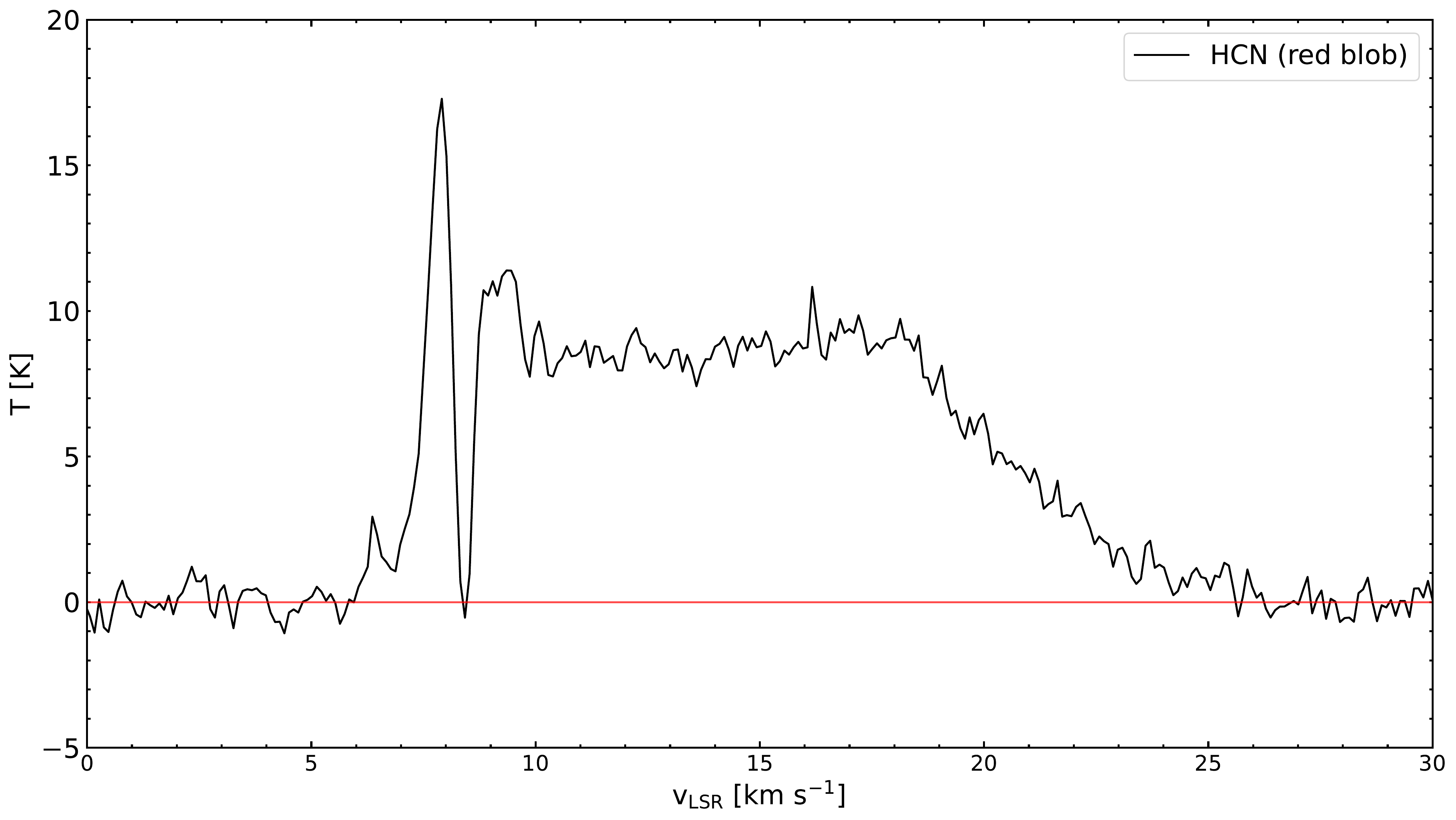}
\caption{
The HCN line spectrum of the red blob from Cycle 3, 2018 data. The spectrum was extracted from an ellipse of dimension 0\farcs6 by 0\farcs3 centered at
19:37:00.611, 07:34:06.05 (J2000).
}
\label{redblob}
\end{figure}

In the south-western cavity wall, there is a bright knot visible in Figure
\ref{CSwalls}. The HCN spectrum taken at this position
(Figure \ref{redblob}) shows a long
tail of emission extending about 17 \kms\ to the red, so we refer to it as 
the ``red blob." This feature is again seen in all three epochs.
None of the other lines show any unusual line profiles at this position.
The position of the red blob is $5\farcs5  = 910$ au SW of the protostar.
An ejection traveling at 100 \kms\ would take about 46 yr to travel this distance.

\section{Models of the Dust Continuum Emission}\label{models}

\subsection{Luminosity Variation}\label{vary1}

\begin{figure}
\center
\includegraphics[scale=0.3, angle=0]{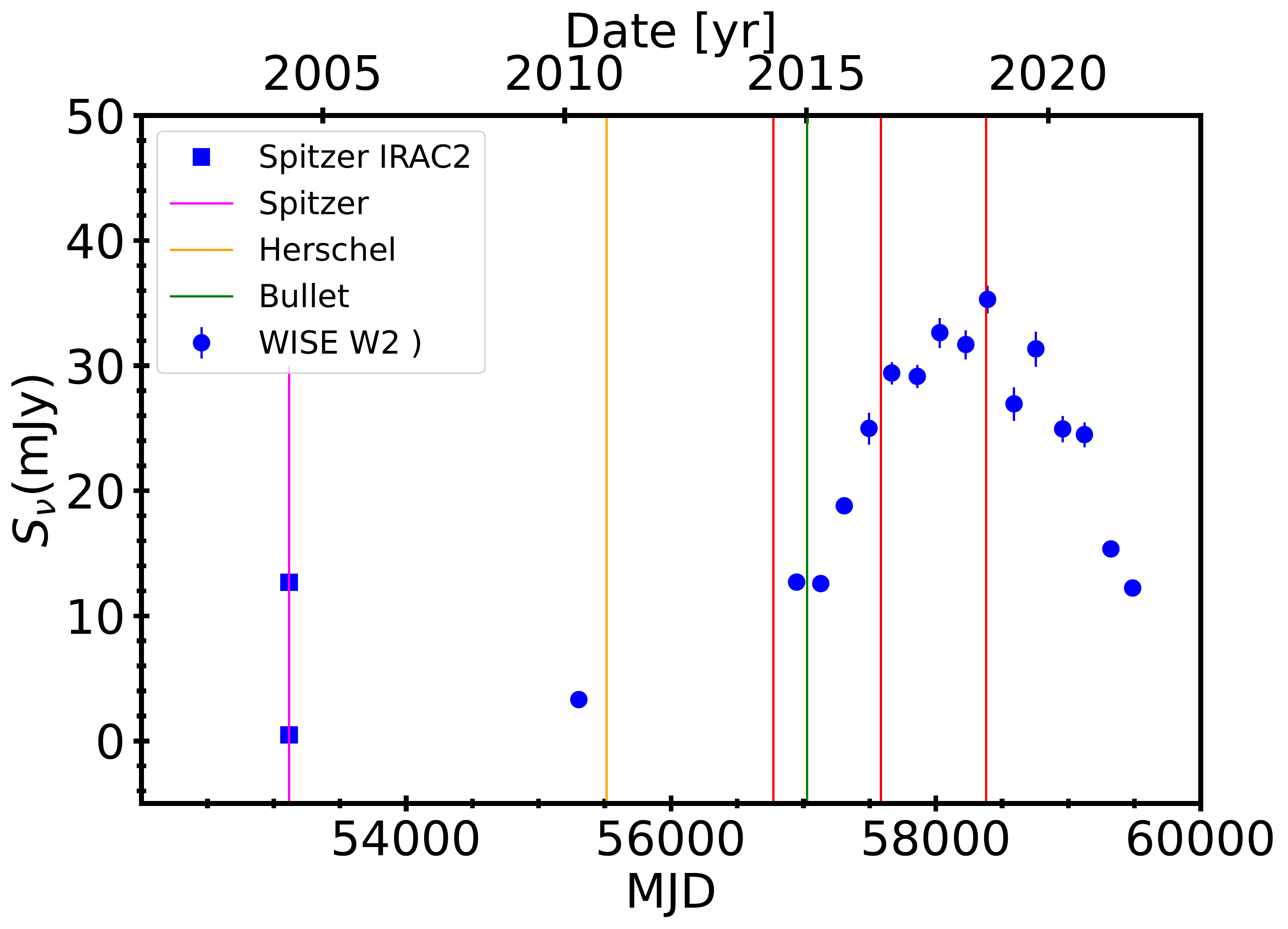}
\caption{
 WISE W2 photometry as a function of MJD (C-H Kim et al. in prep.).
The vertical lines show the dates of various observations. The red lines indicate Band 7 observations from Cycle 1 and two dates for the Cycle 3 data presented here. The dates of the \spitzer\ and \herschel\ observations are indicated by
magenta and orange lines, respectively. 
\added{Photometric points from Spitzer 
\citep{2008ApJ...687..389S}
for 2\farcs4 and 24\arcsec\ apertures are shown with squares.}
The likely earliest date for the ejection of the molecular bullet is indicated by the green line.
}
\label{W2-t}
\end{figure}

The light curve of the W2 (4.5 \micron) flux densities in Figure \ref{W2-t}, along with vertical lines indicating the dates of the \spitzer, \herschel, and selected \alma\ observations, implies that the luminosity of B335 increased substantially after about MJD 56000 and has since declined.
The \spitzer\ data were obtained before the original \wise\ data were available, but the \herschel\ data were obtained close in time to the original \wise\ data point, clearly preceding the luminosity rise. We will assume that the \herschel\ data were also in what we will call the quiescent phase and produce a source model for that phase. Then, we will vary the luminosity in that model to predict the W2 data. After fitting the result, we will have a model for the luminosity as a function of time for the period covered by the W2 data.

\subsection{1D Models}\label{1dmods}

The continuum emission from B335 has been extensively modeled
\citep{2002ApJ...575..337S,2011ApJ...728..143S,2008ApJ...687..389S}
but those predating 2008 assumed a distance of 250 pc.
Spherical models of inside-out collapse 
\citep{1977ApJ...214..488S} are characterized by a sound speed (\cseff)
and an age (\tcol) since the collapse began. 
For practical purposes inner and outer radii of the envelope are
also assumed.
Preliminary versions of these models for the \alma\ data
were discussed by
\citet{2015ApJ...814...22E}.

\begin{figure}
\center
\includegraphics[scale=0.30, angle=0]{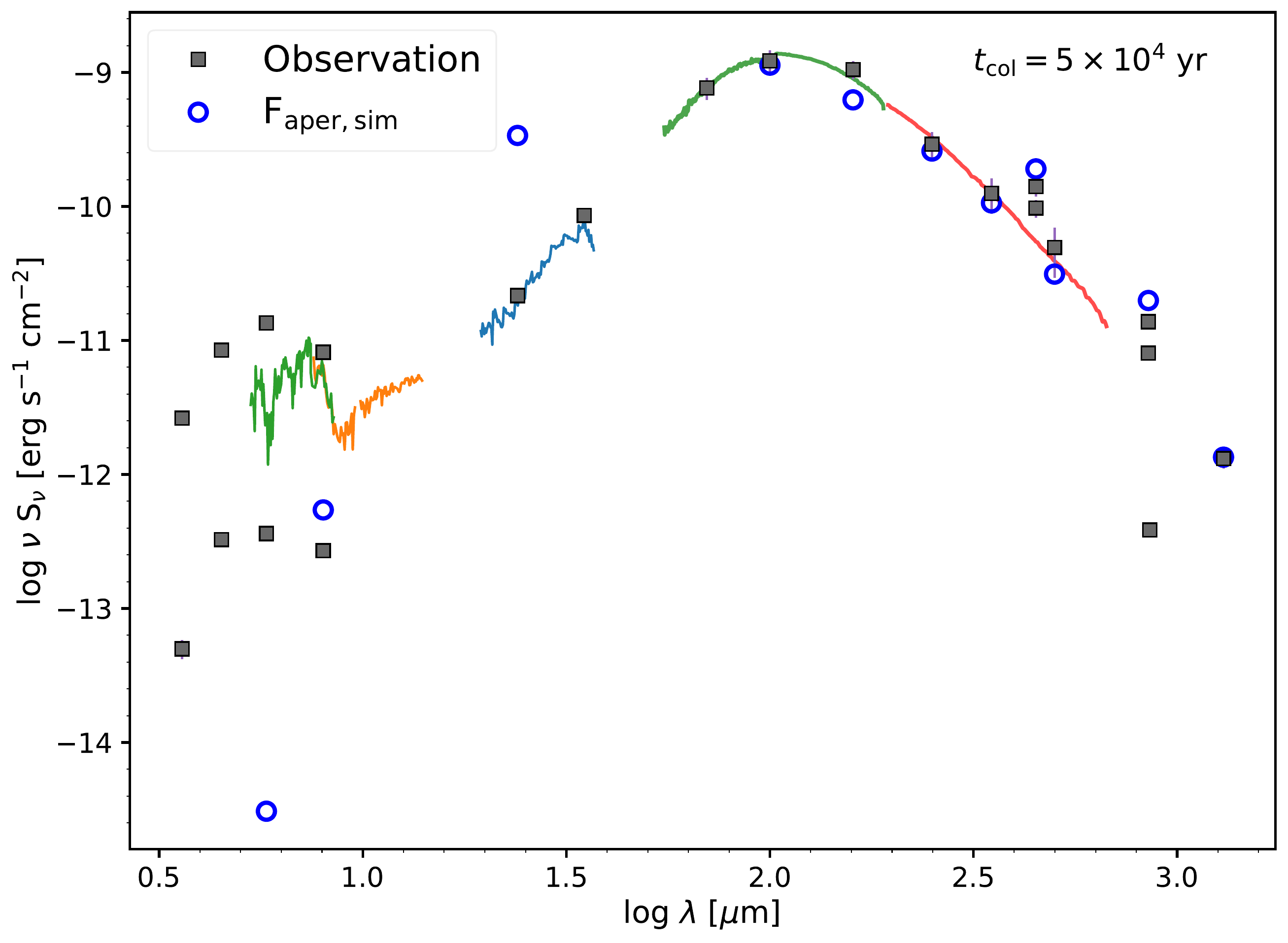}
\caption{
Data and 1D model of the SED for spherical collapse at an age of
$5\ee4$ yr. The data are the same as in Figure \ref{seddata}, and
the blue open circles are the predictions for the large beam
photometry in Table \ref{phottab}. 
}
\label{1Dmod11dust}
\end{figure}

\begin{figure}
\center
\includegraphics[scale=0.35, angle=0]{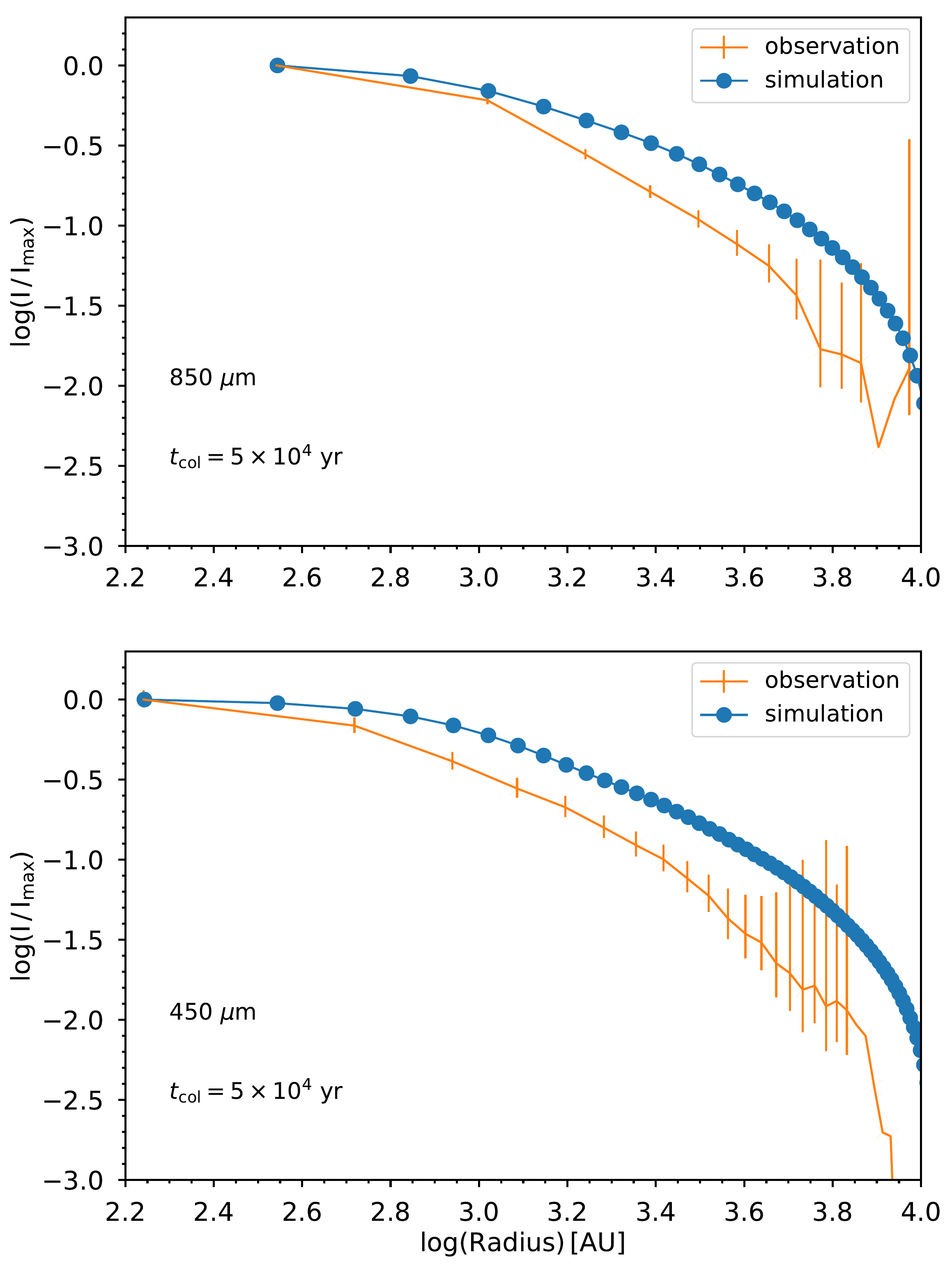}
\caption{
Observed and modeled radial profiles for spherical collapse at an age of
$5\ee4$ yr. The data are plotted with orange errorbars, and
the blue circles are the model predictions. 
}
\label{1Dmod11dustradprof}
\end{figure}

Models of the continuum were compared to photometry (Table \ref{phottab})
and radial profiles at 450 \micron\ and 850 \micron\ emission
\citep{2002ApJ...575..337S}.
These 1D models all had two serious problems. Very young ages
($\tcol < 1\ee4$ yr)
were required to approximate the radial profiles, but these ages
were incompatible with both outflow ages and ages based on modeling
line emission.
Also, 1D models severely underestimate the observed SED at shorter
wavelengths,
 a well-known problem because such models 
have too much extinction at short wavelengths
\citep{2002ApJ...575..337S}.
A 1D model of a spherical envelope for $\tcol = 5\ee4$ yr
 is compared to the observations
in Figure \ref{1Dmod11dust}. 
While the model can reproduce most of the far-infrared and submillimeter
data, it severely underproduces the near-infrared data, and the radial
profiles are far too flat to match the observations (Figure \ref{1Dmod11dustradprof}).

\subsection{3D Models}\label{3dmodels}

More realistic models must consider the three-dimensional structure of
the source, especially the role of the outflow
\citep{2010ApJ...710..470D,  2017ApJ...835..259Y}.
We adopted a model that includes an infalling, rotating envelope, an
embedded disk, and bipolar outflow lobes. 
We use the radiative transport code, \hyperion, an open-source, 
parallelized three-dimensional code 
\citep{2011A&A...536A..79R}.
The convergence criteria for \hyperion\ are applied to the absorption
rate ($\dot A$) of the dust in each cell. The convergence criteria
for most simulations were set so that 95\% of the differences between
iterations were less than a factor of 2, and the 95\% value of the
differences changed by less than 1.02 (about 2\%).


\newcommand{\ra}[1]{\renewcommand{\arraystretch}{#1}}

\begin{table}[htbp!]
  \ra{1.15}
  \caption{Model parameters}
  \centering
  \begin{tabular}{rp{2.5in}}
    \toprule
    \multicolumn{2}{c}{\bf Envelope parameters} \\
    \hline
    \tcol\ & Age of the protostellar system after the start of collapse. \\
    $\cseff$ & Effective sound speed of initial model. \\
    $\omegaz^{\rm a}$ & The initial angular speed of the cloud. \\
    $\router^{\rm a}$ & Outer radius of the envelope  \\
    \multicolumn{2}{c}{\bf Disk parameters} \\
    \hline
    $\mdisk$ & Total mass of the disk. \\
    $\beta$ & The flaring power of the disk. \\
    $\diskheight$ & The disk scale height at 100~au. \\
    \multicolumn{2}{c}{\bf Outflow cavity parameters} \\
    \hline
    $\thetacav^{\rm a}$ & The cavity opening angle  \\
    $\rhocav$ & The dust density of the inner cavity. \\
    $\rcav$ & The radius where the cavity density starts to decrease. \\
    $\thetaincl^{\rm a}$ & The inclination angle of the protostar, 0$^{\circ}$ for face-on and 90$^{\circ}$ for edge-on view. \\
    \multicolumn{2}{c}{\bf Stellar parameters} \\
    \hline
    $\tstar^{\rm a}$ & The temperature of the central protostellar source assuming blackbody radiation. \\
    $\rstar$ & The radius of the central protostellar source. \\
    \bottomrule
    \multicolumn{2}{p{3in}}{$^{\rm a}$These parameters are fixed in the search of the best fit model.} \\
  \end{tabular}
  \label{model_params}
\end{table}

Our modeling approach follows that of 
\citet{2017ApJ...835..259Y},
who explained the basic method and explored the effect of the different
parameters on the observed SED. We summarize the input model 
construction here.
All the parameters are summarized in Table \ref{model_params}.
The envelope is a rotating, infalling envelope that was originally
spherical and isothermal
\citep{1984ApJ...286..529T},
the TSC model.
This model is defined primarily by the effective
sound speed (\cseff), the age since collapse began (\tcol),
and the initial angular velocity (\omegaz).
In addition, we specify the  outer radius for the
envelope (\router). The inner radius (\rinner) 
is set by the dust evaporation temperature in an iterative way.
The physical model is axisymmetric, so 2D, but the inclination angle
adds a dimension, so \replaced{we refer to these models as 3D}{the radiative transfer is done in 3D}.

Because of the observational constraints on disk size 
\citep{2011ApJ...742...57Y,2015ApJ...799..193Y},
the disk is not very significant in producing the observed SED, but
we model it as a flared disk with parameters described by
\citet{2017ApJ...835..259Y}.
The disk is specified by its mass (\mdisk), flaring power-law ($\beta$), 
and scale height at 100 au (\diskheight). These parameters do affect
the near-infrared ($3.6 - 8$ \micron) part of the SED.

We do not model the outflow itself, but only the effect of the cavity
on the SED and radial profiles.
The outflow lobes are characterized by their shape and the density
structure inside them. The shape is defined in cylindrical coordinates,
($z$ and $\varpi$), where  $z$ expresses the distance along
the outflow axis and $\varpi$ measures the perpendicular distance from that
axis to the edge of the lobe. The equation of the surface is
$z = c_0 \varpi^{1.5}$, where
\begin{equation}
c_0 = \frac{\cos\thetacav}{r^{0.5}_{\rm fid} \sin^{1.5}\thetacav},
\end{equation}
and \thetacav\ is the half-angle of the outflow lobe measured
at $ r_{\rm fid} = \eten4$ au,
where $ r^2 = \varpi^2 + z^2$.
This model produces a curved shape that is qualitatively similar
to the shape seen faintly in the Cycle 1 \alma\ continuum data
\citep{2015ApJ...814...22E}
and in Figure \ref{CSwalls}.
The mass inside the outflow cavity is characterized by a small
region of constant density (\rhocav) that extends out to a radius (\rcav) beyond
which the density declines as a power-law ($\rho \propto r^{-2}$).
The axis of the outflow is assumed to be aligned with the rotation axis.
Finally, the inclination angle of the rotation and outflow axes relative
to the line of sight (\thetaincl) has to be specified. A face-on rotation
axis has $\thetaincl = 0$. 

In \hyperion\, 
the luminosity source 
is characterized by a stellar radius (\rstar ) and temperature (\tstar ).
For deeply embedded sources, these are nearly degenerate, and only
the stellar luminosity (\lstar) is relevant. We kept \tstar\ fixed and
varied \rstar\ to change the luminosity. The actual luminosity arises from
accretion, so these parameters are not meant to be realistic. The value
of \tstar\ affects the near-infrared portion of the SED somewhat.

The standard dust model is that of column 5 of table 1 of 
\citet{1994A&A...291..943O}
as extended to shorter wavelengths with anisotropic scattering by
\citet{2005ApJ...627..293Y}.
The standard Henyey-Greenstein model 
\citep{1941ApJ....93...70H}
for the angular distribution
of scattering is used, unless otherwise noted.


\begin{table}[htbp!]
  \ra{1.15}
  \caption{Best-Fit Model parameters}
  \centering
   \begin{tabular}{l c c}
    \toprule
    \multicolumn{3}{c}{{\bf Envelope parameters}} \\
    \hline
    $\tcol$ & 40000 years & \\
    $\cseff$ & $0.30~{\rm km~s}^{-1}$ &  \\
    $\omegaz$ & $2.0 \times 10^{-14}~{\rm s}^{-1}$ & (fixed) \\
    $\router$ & $2\ee4$~au & (fixed) \\
    \multicolumn{3}{c}{\bf Disk parameters} \\
    \hline
    $\mdisk$ & $0.063 $    & (fixed) \\
    $\beta$ & 1.3 &  \\
    $\diskheight$ & 4.0 AU &  \\
    \multicolumn{3}{c}{\bf Outflow cavity parameters} \\
    \hline
    $\thetacav$ & $27.5^{\circ}$ & (fixed) \\
    $\rhocav$ & $4 \times 10^{-20}~\rm {g~cm}^{-3}$ & \\
    $\rcav$ & 20 AU & \\
    $\thetaincl$ & $87^{\circ}$  & (fixed)  \\
    \multicolumn{3}{c}{\bf Stellar parameters} \\
    \hline
    $\tstar$ & 7000 K & (fixed) \\
    $\rstar$ & 1.17 $\rm R_{\rm \odot}$ &  \\
    \bottomrule
  \end{tabular}
  \label{best_fit}
\end{table}

Given the very large number of parameters in a 3D model, it is necessary
to constrain as many as possible from existing information.
The values of parameters that were fixed in the models are listed in
Table \ref{best_fit} with a note that they were fixed. 
The distance was fixed at 164.5 pc
\citep{2020RNAAS...4...88W}.
The initial angular velocity, $\Omega_0$, is set at 2\ee{-14}
s$^{-1}$, based on the arguments in
\citet{2019ApJ...871..243Y},
as noted in \S \ref{intro}.
\added{The outer radius was estimated from Herschel maps by
\citet{2013A&A...551A..98L} 
to be 4\ee4 au, adjusted to the new distance.}
The outer radius, \rout, is set at 2\ee4 au, \replaced{which is arbitrary,}{about half the estimate from Herschel,}
but larger radii did not change the results.
The mass enclosed in that value
of \rout\ is 3.3 \msun.
The opening angle (\thetacav) was set to 27\fdg5, based on
analysis of the CO outflow in the context of detailed
source models
\citep{2011ApJ...728..143S}
and CO maps by
\citet{2008ApJ...687..389S}. 
Most models assumed $\thetaincl = 87\degree$,
based on best fits to the SED by
\citet{2008ApJ...687..389S}
using model grids and a fitter
\citep{2006ApJS..167..256R, 2007ApJS..169..328R}.

The values of other parameters are the result of
the optimization of the predicted SED and the radial profiles.
As discussed at length in the appendix of
\citet{2017ApJ...835..259Y},
the radial profiles and different regions of the SED best constrain
different parameters.
To compare the model predictions to the observations, a simulated
SED was produced by calculating an emission image at a particular
wavelength, using the ray-tracing part of \hyperion, and convolving
that image with a top hat aperture of the specified size. Azimuthally
averaged radial profiles were also computed from the model after 
simulating the beam used for the observations. Images of the model at
specific wavelengths can also provide constraints. We will describe
the steps used to optimize the model parameters in the order that they
were taken, but the final optimization involves iteration between correlated
parameters.

First the luminosity has to be constrained. In the context of these
models, a blackbody star is assumed. The stellar parameters of temperature (\tstar) and radius (\rstar) enter in combination 
($ L \propto \tstar^4 \rstar^2$).
To match the {\it observed} luminosity of 1.36 \lsun\ with 
$\thetaincl = 87\degree$, the central
luminosity must be 2.95 \lsun,  2.2 times
the observed luminosity. This result is a feature of \replaced{3D models}{models with outflow cavities},
which \replaced{account for the tendency of}{cause the} radiation to be channeled 
through the outflow lobes, and the effect is large in B335 because the outflow
is nearly edge-on.

Second, the total mass of the envelope is best
constrained by the longest wavelength photometry, given a model
of dust opacities. In Shu-type models, this is set by the sound
speed (\cseff) because the density depends on the sound speed:
$ \rho = \cseff^2/(2 \pi G r^2)$. 
The new, larger distance necessitated some changes in parameters used in
recent modeling
\citep{2015ApJ...814...22E}.
The effective sound speed, including turbulence, $\cseff$ had been
taken to be $0.233$ \kms\
\citep{2015ApJ...814...22E}
in models of the source at 100 pc distance.
At the larger distance, a larger sound speed is needed to match the flux
into large beams at the longest wavelengths, with a value of $\cseff = 0.30$ \kms\ working well.
The previous value was obtained by adding in quadrature the thermal sound
speed for an initial $\tk = 13$K 
\citep{2005ApJ...626..919E}
 and the microturbulent contribution
inferred from line widths in single-dish beams
\citep{2015ApJ...814...22E}. 
The obvious justification for increasing \cseff\ is that the Alfven speed has 
not yet been included. If added in quadrature
\citep{1987ARA&A..25...23S}, 
 the required Alfven speed would be $0.19$ \kms. The Alfven speed is $B(4 \pi \rho)^{-0.5}$.
The large scale magnetic field is estimated at $30.2\pm17.7$ $\mu$G
\citep{2020ApJ...891...55K}
but the relevant, pre-collapse density is unclear in the presence of a
density gradient. The required Alfven speed can be obtained if the relevant
density is about $5\ee4$ \cmv, a very reasonable value.
\citet{2021A&A...645A.125Z} derived a field value of $142\pm46$ $\mu$G from
observations at 214 \micron\ and an assumed density of $\rho = 5.41\ee{-18}$
g \cmv. This combination leads to nearly the same Alfven speed of $0.17$
\kms.
The change in \cseff\ leads to changes in many other derived properties
compared to previous models
\citep{2015ApJ...814...22E}. 
These will be summarized in \S \ref{sourceprop} after the other modeling parameters are
discussed.

\begin{figure}
\center
\includegraphics[scale=0.35, angle=0]{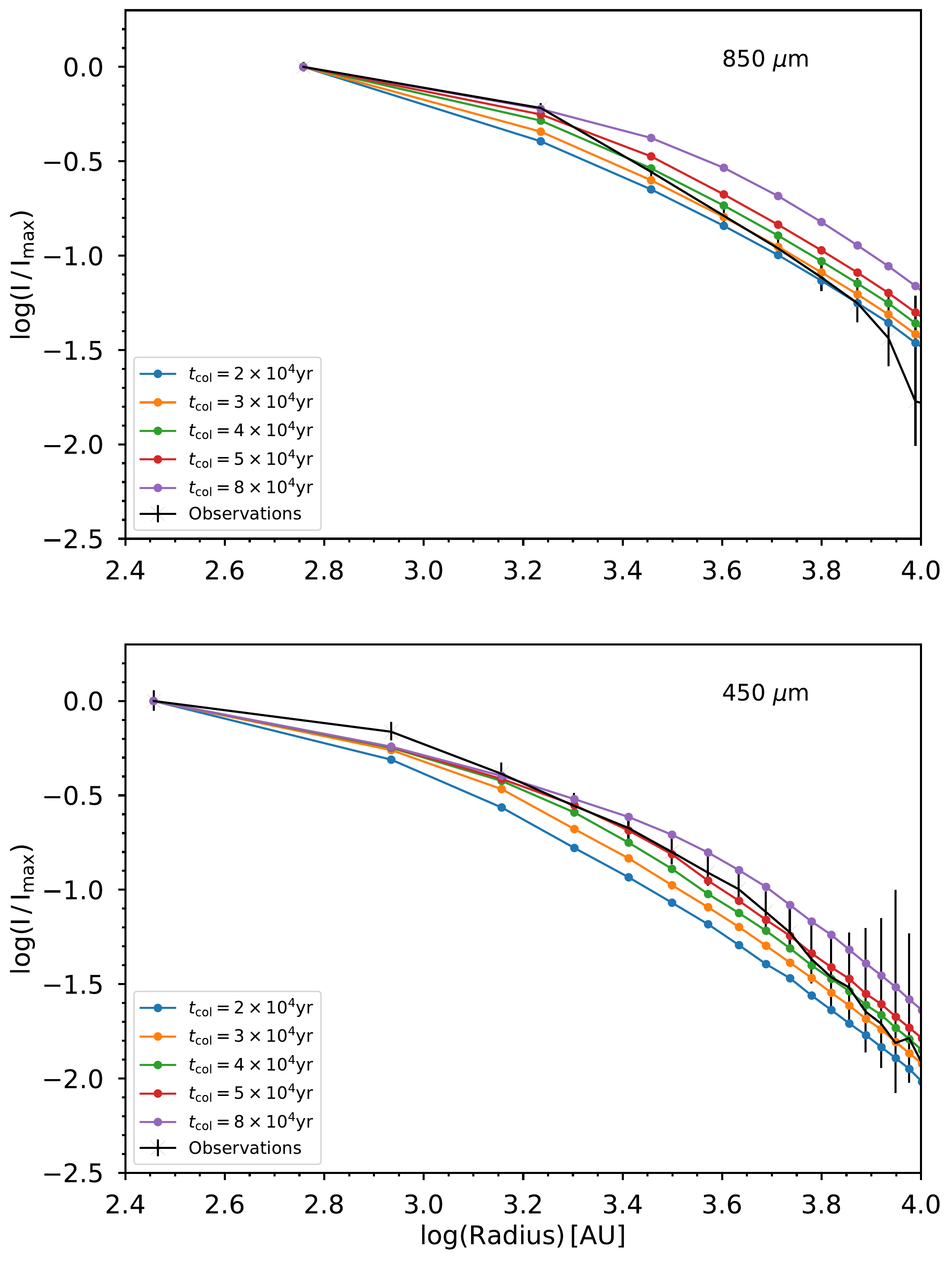}
\caption{
Observations and 3D model of the radial profile of the submillimeter wavelength 
emission. Models with $\tcol = 3\ee4$ yr, $\tcol = 4\ee4$ yr,
 $\tcol = 5\ee4$ yr, and $\tcol = 8\ee4$ yr are  shown. 
}
\label{agetest}
\end{figure}

Third, the age of the source must be constrained. In TSC models, the age
is the time elapsed since the initiation of infall (\tcol). The age and 
sound speed determine the infall radius ($\rinf = \cseff \tcol$), 
which determines
the shape of the density profile. In turn, the density profile is the main determinant of the
normalized, azimuthally averaged, radial profile of emission at submillimeter
wavelengths (i.e., the radial profile). 
As the source evolves, the initial $\rho \propto r^{-2}$ density profile
shifts to a flatter profile, leading to a flatter predicted radial intensity
profile (Fig. 25 of \citealt{2017ApJ...835..259Y}).
Since we have adjusted \cseff\ in the second
step, we vary \tcol\ to best approximate the observed radial intensity profiles.
The observed radial profiles are annular averages, so we produce the model
radial profiles in the same way, thus accounting for the effect of the outflow cavity.
Models with different \tcol\ are compared to the observations in
Figure \ref{agetest}. The radial profile at 450 \micron\ is best matched
at $\tcol = 5\ee4$ yr, but the profile at 850 \micron, which is less sensitive
to the temperature distribution, favors $\tcol = 3\ee4$ yr. 
An age of $4\ee4$ yr predicts a radial profile at 850 \micron\ that is slightly
too flat and one at 450 \micron\ that is slightly too steep, thus
presenting about the best compromise.
The larger distance and larger \cseff\
partially cancel out in their effects on the radial intensity profiles of
the model, so that the best-fit $\tcol = 4\ee4$ yr is not
very different from the age of 5\ee4 yr favored by
\citet{2015ApJ...814...22E}. 
We adjusted \mdisk\ to be 0.25 of 
the collapsed mass ($\mstar + \mdisk$) because larger ratios tend to be unstable. For the age and infall rate fixed
by the above considerations, $\mstar = 0.25\ \msun$, so $\mdisk = 0.063\ \msun$.
For comparison, \citet{2015ApJ...814...22E} set a limit on the optically thin mass based on
the Cycle 1 data; after updating to the new distance, the limit would be about
2\ee{-3} \msun. However, our modeling shows that much larger disk masses can be hidden by optically thick 
dust in the small disk of our models. 

Fourth, the mid-infrared part of the SED is most affected by the
cavity properties, especially the density at the base of the 
cavity, with some effects from the disk properties.
The best match was found for a radius for the region of 
constant density in the cavity (\rcav) of 20 au, 
and the density of the constant region (\rhocav)
of 4\ee{-20} g \cmv.
However, these parameters also affect the SED at shorter wavelengths
because the flux densities at those wavelengths are dominated by light scattered
from the outflow cavity. The density that produces enough emission at 24 \micron\ somewhat overproduces the emission in the small apertures in the 
IRAC bands, 
unless the disk and star parameters can compensate.

The near-infrared part of the SED (the IRAC bands) depends on many parameters,
including the rotation rate, inclination angle, cavity properties,
disk properties, and stellar properties. Deviations from the assumed simple, 
azimuthally symmetric structure can also have a big effect, so
a good fit to the IRAC flux densities cannot necessarily be interpreted as actually representing
the structure. With that caveat in mind, we describe the adjustments
used to provide a decent match to the near-infrared observations.
A lower \tstar, with the larger \rstar\ needed to maintain the luminosity,
produced more emission in the near-infrared. A better match was found for
the (rather unrealistic) value of $\tstar = 7000$ K. 
We produced small grids of \omegaz, \thetaincl, $\beta$ and \diskheight\ to 
optimize the near-infrared SED. The resulting best-fitting
 parameters are listed in Table
\ref{best_fit}. The common feature of these best parameters is that they
minimize the emission in the IRAC bands. Some of the parameters are not
very realistic. \tstar\ is quite large for a star of 0.25 \msun, and the 
disk parameters ($\beta$, \diskheight) produce a very flat disk. Even with 
these rather extreme parameters, the IRAC fluxes into small apertures
are over-produced, especially at 3.6 \micron. More typical stellar and
disk parameters could be used if there is an extra source of extinction
near the source, as will be discussed below.

\begin{figure}
\center
\includegraphics[scale=0.3, angle=0]{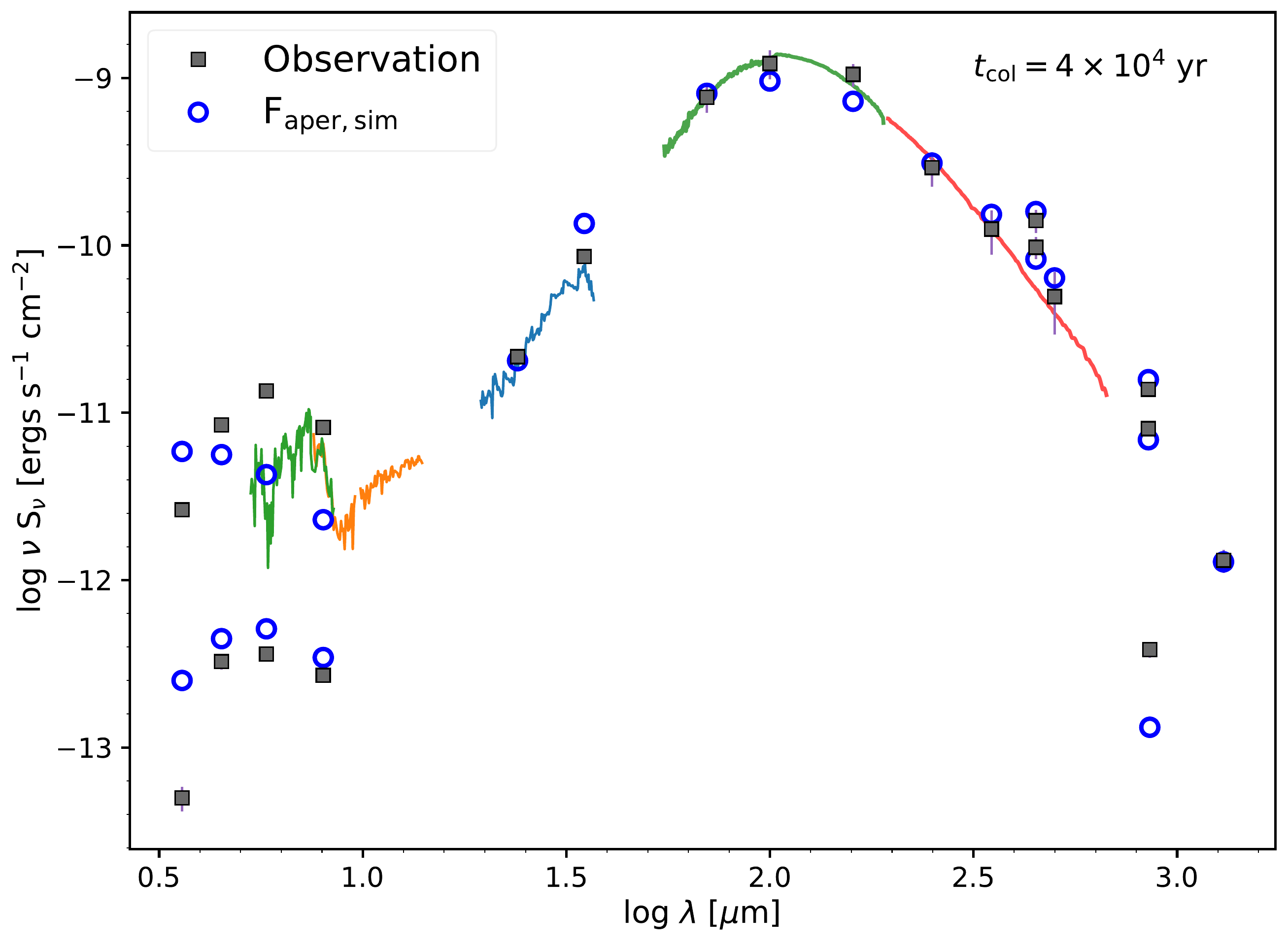}
\caption{
Observations and best 3D model of the continuum SED, 
with the parameters in Table \ref{best_fit}. 
}
\label{3dbestdust}
\end{figure}

The SED of the best model for the continuum is shown in Figure \ref{3dbestdust}.
The SED is well matched except in the near-infrared, where the
model over-predicts the emission into small beams and under-predicts
the emission into large beams, and the ALMA observations, which we discuss below.  At 3.6 \micron, the model strongly over-predicts the emission into both large and small beams. Variations in the
disk, cavity, and stellar properties were used to optimize the
near-infrared, but the over-prediction into small beams was a robust result. 
The 3D models produce a much better (though still
imperfect) match to the short wavelength emission than did the 1D model
with the same age.  The outflow cavity has clearly allowed more radiation
at mid-infrared and near-infrared wavelengths to reach the observer.
The emission is essentially all scattered by the outflow cavities into the line of sight.

\begin{figure}
\center
\includegraphics[scale=0.35, angle=0]{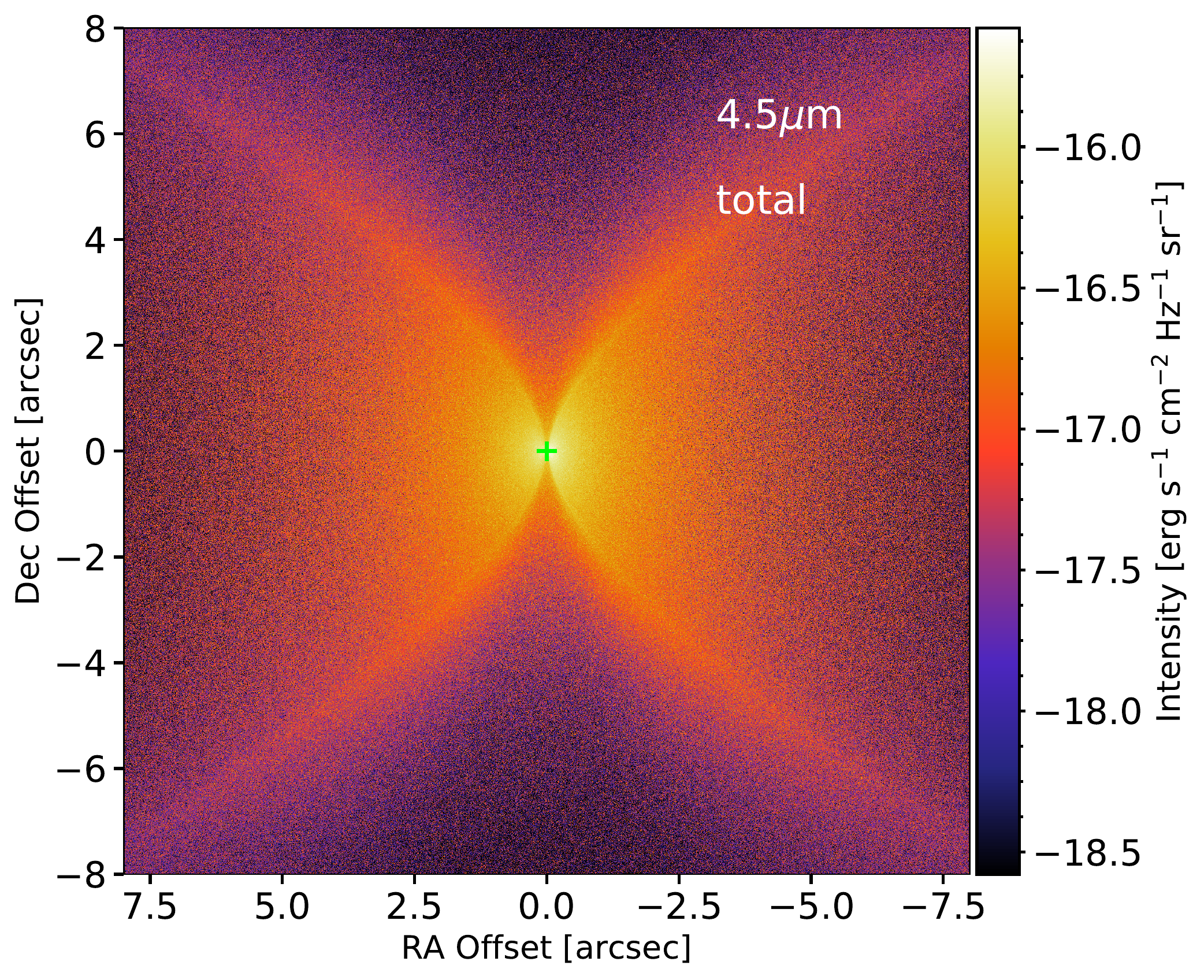}
\caption{
Image at 4.5 \micron\ based on the continuum model with parameters
in Table \ref{best_fit}. 
\added{The scale is adjusted to match that from Figure \ref{CSwalls}}. 
}
\label{imageirac2}
\end{figure}

Figure \ref{imageirac2} shows a simulated image at 4.5 \micron; 
two fan-shaped nebulosities, strongly peaked near the source and brighter
along the cavity wall, are clearly seen. The shape agrees \added{reasonably} well with the
image of the cavity walls indicated by the CS emission (Figure \ref{CSwalls}) 
and roughly with
the image in Figure 9 of 
\citet{2008ApJ...687..389S}, but the model is more symmetric between the
two lobes than are the infrared observations.

\begin{figure}
\center
\includegraphics[scale=0.4, angle=0]{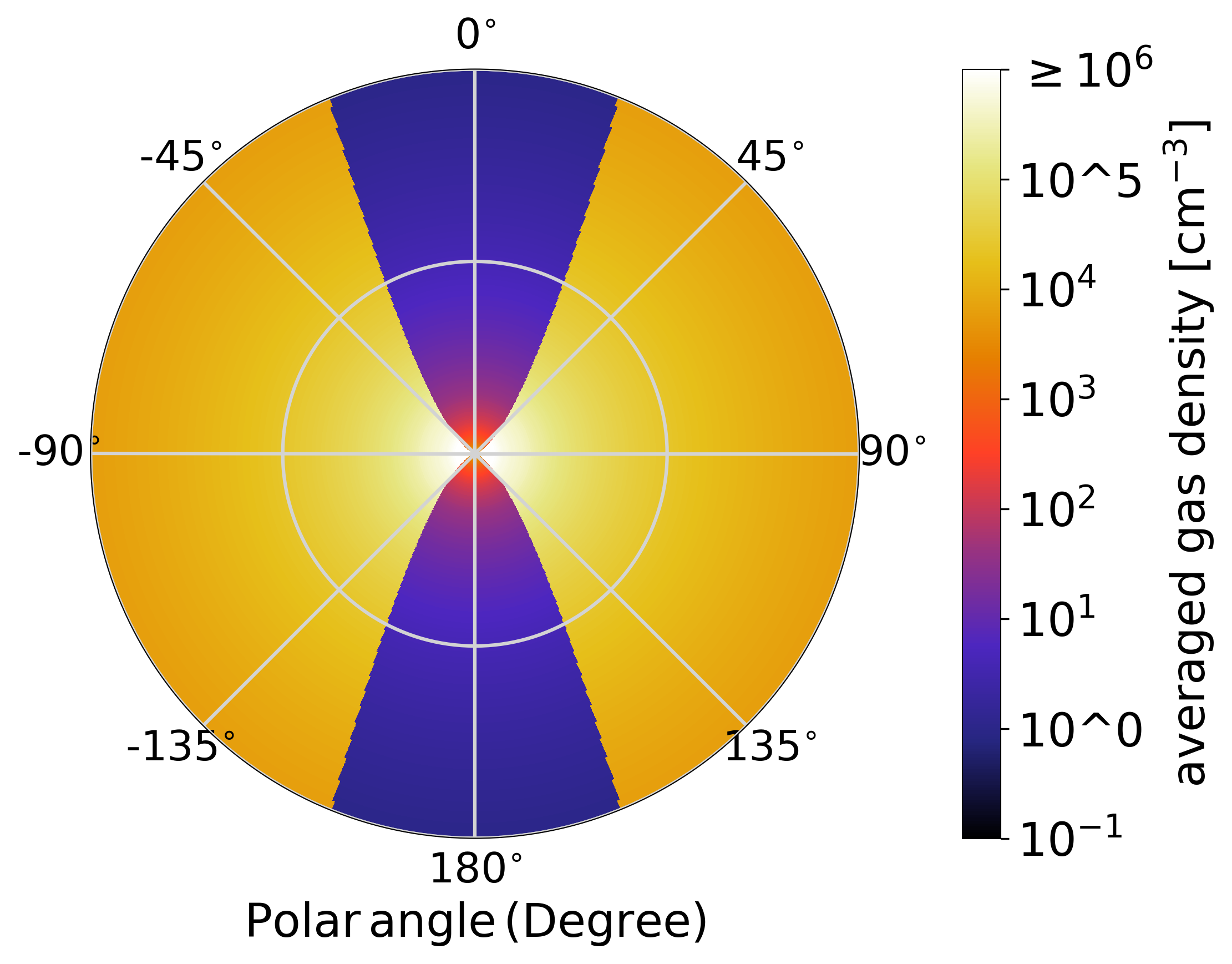}
\caption{
The  distribution of average gas density as a function of radius and 
polar angle. Note that the model polar angle is taken relative to the z axis.
For comparison to data, we rotate the model predictions to the E-W outflow
direction.
}
\label{densityimage}
\end{figure}

\begin{figure}
\center
\includegraphics[scale=0.3, angle=0]{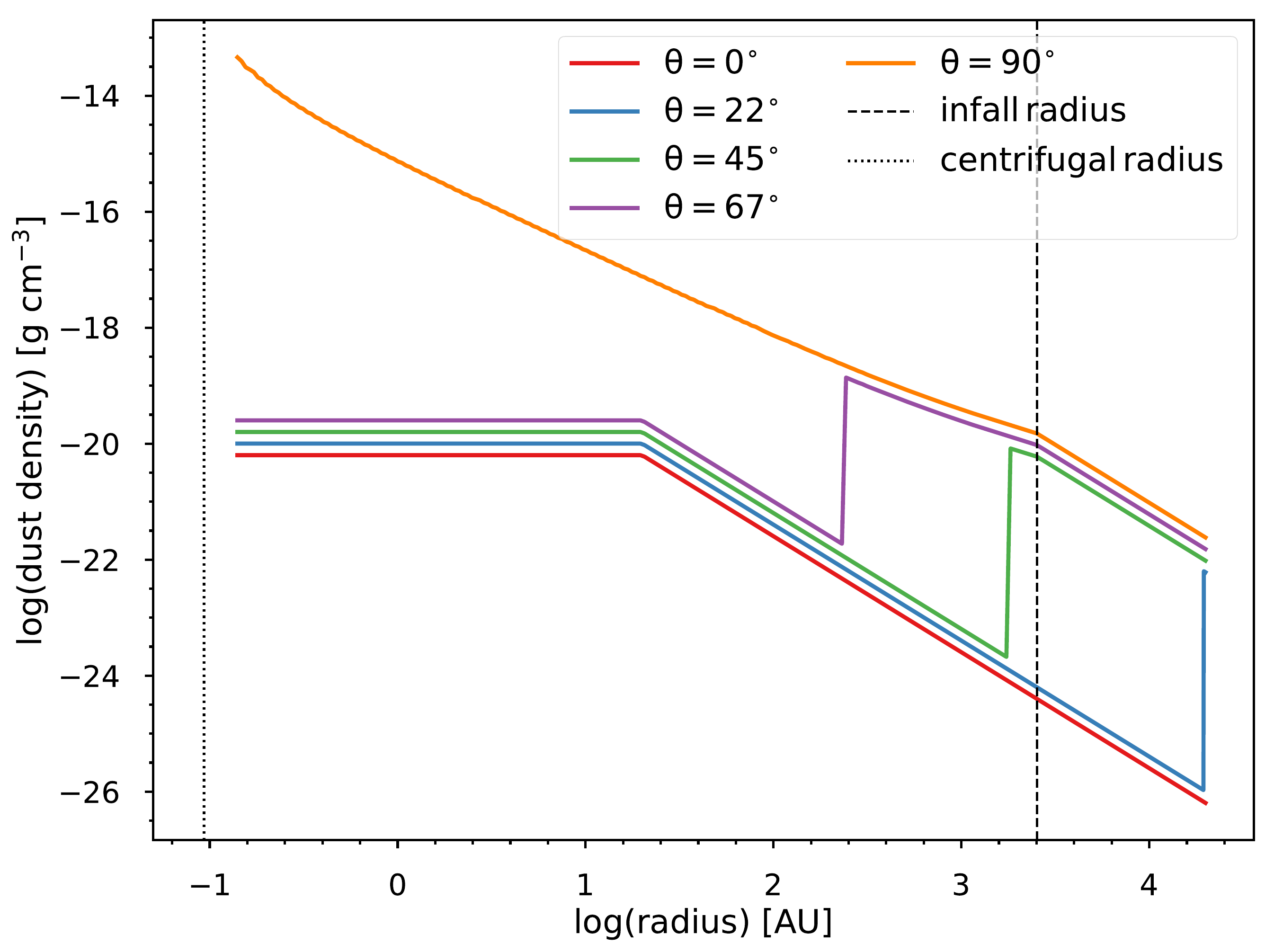}
\caption{
The radial distribution of dust density at five different polar angles, as
defined in Figure \ref{densityimage}. Except for $\theta = 90\degree$, the profiles
start in the cavity and increase sharply at the radius where the line-of-sight 
passes out of the cavity into the envelope.
}
\label{radialdensity}
\end{figure}

The density structure of the best-fitting model is shown in Figure
\ref{densityimage}, which also defines the polar angle of the model,
$\theta$. The model has the outflow cavities along the z axis by convention,
so this image is rotated by 90 degrees from the one in Figure
\ref{imageirac2}.
The radial distributions of dust density along five distinct polar angles are
shown in Figure \ref{radialdensity}. The sharp increases show when the
line of sight passes from the cavity into the envelope.

One glaring failure of the continuum model is the weakness of the emission
predicted for the \alma\ emission into a 2\arcsec\ aperture. The best model predicts a flux that is a factor of 3 less than the Cycle 1 observations.
Even the Cycle 1 observations may have occurred after the beginning of the
rise in luminosity, and the continuum flux increases by a factor of 2 by the
time of the 2018 observations. However, models with higher luminosity also
fail to reproduce the \alma\ data at any of the three epochs.
A wide range of parameters for the disk,
\omegaz, etc. were tried, but the predicted emission into a 2\arcsec\ aperture
was rather insensitive to these variations and was never close to the observed emission.

To compare to the \alma\ data in more detail, a model image was made at the mean
wavelength of the \alma\ data and converted into uv-plane data using the CASA task simobserve for the Cycle 1 data. Then simanalyze was used to
create an \alma\ image which was fitted
with an elliptical Gaussian within a 2\arcsec\ circle. 
The emission in the model has a lower peak intensity and 
is much more elongated than the observed emission,
being defined by the outflow cavity walls. 
So, neither the shape nor the flux are reproduced well. 
A similar issue was seen in BHR71 by 
\citet{2020ApJ...891...61Y},
who suggested the presence of 
an additional structure between the envelope and the Keplerian
disk. 
The \alma\ continuum data strongly suggest that there is a structure on the
scales of $\sim 30$ au in addition to the envelope and disk in our models.
Such a structure could also provide additional extinction at short wavelengths,
as suggested above in the discussion of the discrepancies in the near-infrared.
The most likely candidate is something like a pseudo-disk, but a thermal 
radio jet could also contribute. 
\citet{2016MNRAS.460.1039P} found spectral indices as high as 1.6 in their
study of radio jets from 5 to 23 GHz. If such a large index persisted to 345 GHz, the largest flux found by
\citet{2004RMxAA..40...31G}
could account for all the continuum that we observe. For a more typical index
of 0.3, the contribution would be negligible.

In summary, 3D models with outflow cavities can reproduce observations
of both the radial profile and most of the SED with TSC models of reasonable ages
($\tcol = 3 - 5\ee4$ yr). The outflow cavities are the critical element
in resolving issues found in 1D models.

\subsection{Luminosity Variation}\label{lumvary}

\begin{figure}
\center
\includegraphics[scale=0.3, angle=0]{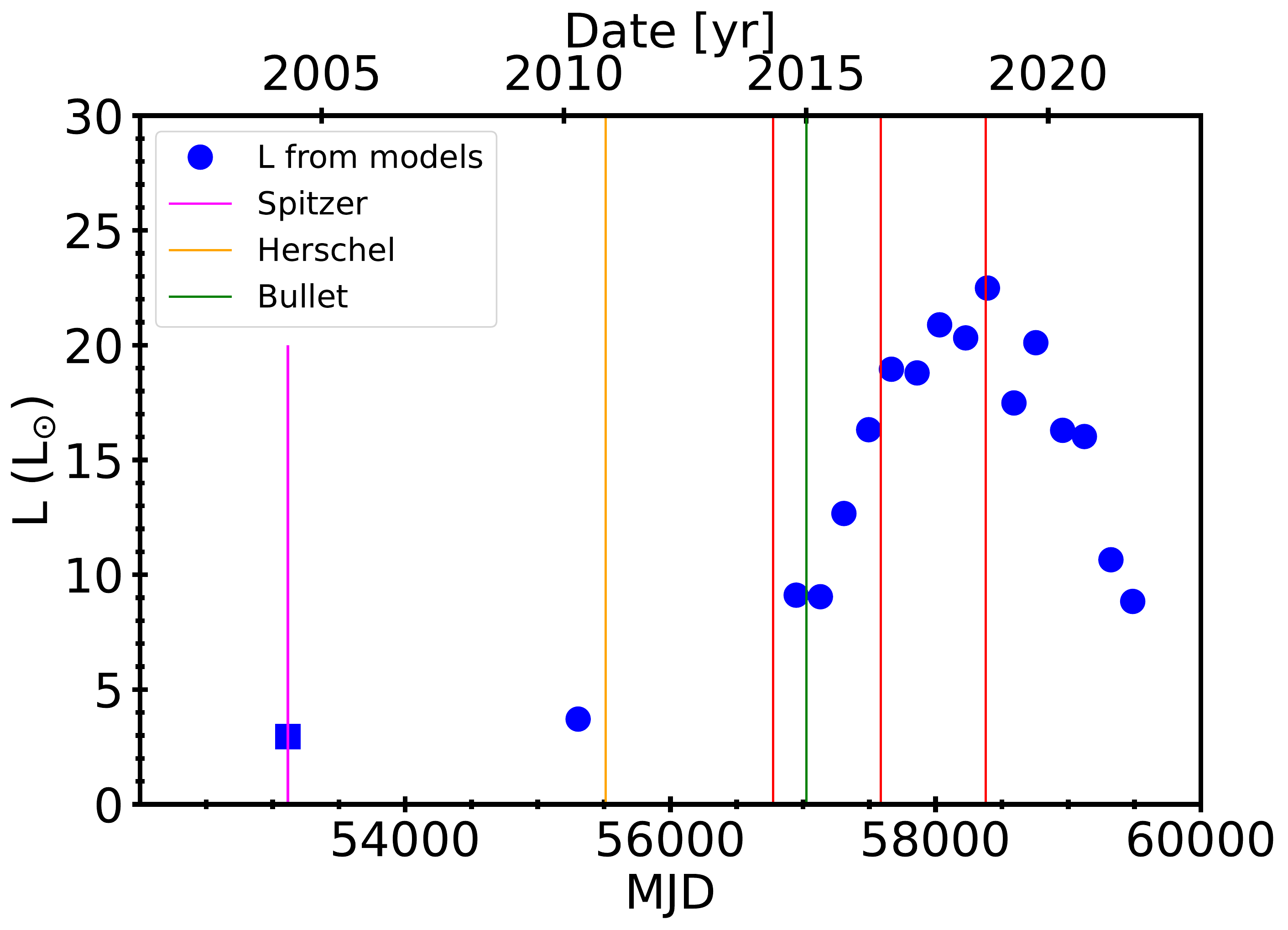}
\caption{
The central luminosity as a function of time based on modeling the
WISE W2 photometry. The vertical lines are the same as in Figure \ref{W2-t}.
\added{The quiescent luminosity is plotted at the time of the Spitzer observations.}
}
\label{lumvst}
\end{figure}

With a model for the quiescent source, we can relate the W2 photometric variation to variations in the luminosity. We ran a series of models, changing only the stellar radius (a stand-in for central luminosity), each predicting the W2 photometry. Then the relation between the model luminosity and the W2 flux density from the model was fitted with a quadratic function between about 3 \lsun\ and 25 \lsun.
The best-fitting model was
\begin{equation}
L/\lsun = 1.837 + 0.845 S_{\nu}(W2) + 0.001 S_{\nu}(W2)^2
\end{equation}
with $S_{\nu}(W2)$ the \wise\ photometry in mJy.
The actual photometry as a function of time, multiplied by a factor of 0.67
to account for a slight under-prediction by the models, then allows
a model for the central luminosity as a function of time  (Figure \ref{lumvst}). 
This model predicts a luminosity increase of a factor of 5-7 over the quiescent state.

In principle, the W1 photometry might be preferred because the W2 bandpass 
includes likely strong line emission from \hh\ and CO 
\citep{2022ApJ...929...60Y},
\added{perhaps balanced by absorption by CO, CO$_2$, and other ices}.
However, our model for the W2 emission matches the observations better than it does for the W1 bandpass. Time sequence far-infrared photometry is what is really needed. There is one useful data point at 214 \micron\ for MJD 58044, near the peak luminosity
\citep{2021A&A...645A.125Z}.
Our luminosity model predicts $L \approx 20$ \lsun\ at that date, but the 214 \micron\ flux density is better matched in a model with $ L \approx 6$ \lsun.
There will be a time delay before the luminosity burst covers the full
region that contributes to the far-infrared emission into relatively large
beams
\citep{2022ApJ...937...29F}, 
which would imply a luminosity larger than 6 \lsun\ at the time of the
214 \micron\ observations.
Consequently, the actual increase is rather uncertain, but likely a factor between 5 and 7.

\section{Models of Line Emission}\label{linemodels}

\subsection{3D Models}\label{3Dlinemodels}

To see how 3D models with outflow cavities and the small amount of 
rotation seen in B335
would fare, we used LIME
\citep{2010A&A...523A..25B}
to model the line emission from the models
computed for the dust emission with \hyperion. Because our dust model
underestimates the ALMA observations, 
we also used a custom code (LIME-AID) to add a continuum source at the center 
of the model equal to the observed flux density
\citep{2020ApJ...891...61Y}.
We used $5\ee4$ intensity grid points (pIntensity) for running models,
but doubled that to $1\ee5$ for the final models. Using twice the number
produced slightly smoother spectra, but the differences were very minor.

The spectral line models used the density and
velocity fields from the TSC models, supplemented by the
cavities added to model the continuum emission. In the inner regions, the 
gas temperature was assumed equal to the dust temperature from the 
\hyperion\ models, including the cavities, so the temperature is higher 
around the cavities. 
For radii larger than 2600 au, the gas temperature was increased,
based on the models including gas heating by photoelectric heating
by the interstellar radiation field (Fig. 8 of
\citealt{2005ApJ...626..919E}).
\added{Increased temperature on the outside of isolated globules is supported by the models of Herschel data
\citep{2013A&A...551A..98L}.}
Because the adaptive gridding used by LIME was unduly distorted
by the sharp discontinuity of the outflow cavity, we did not
use the particle volume density inside the cavity of the continuum 
models, but instead set the abundance 
of the species being modeled to zero inside the cavity. The higher
temperatures near the cavity were captured from the continuum models, but there
were no changes in the gas velocity, so the outflows were not modeled.

The new variables introduced for the
lines are those that characterize the microturbulence
and the abundance profile. 
The microturbulence was taken as
$b = 0.12$ \kms, as in previous models, and assumed to be
independent of radius. 
The rotation was already in the TSC models, but 
for the line simulation, the sense of rotation matters as well. 
Based on a map of H$^{13}$CO$^+$ \jj10\ emission,
\citet{2013ApJ...765...85K}
found a velocity gradient from south to north, indicating
that the north side is rotating away from us on large scales
($r \sim 2\ee4$ au).
\added{The opposite sense of rotation was inferred from the CO 
\jj21\ maps by
\citet{2008ApJ...687..389S}.
}
Because the rotation in the models is
clockwise as viewed from the ``north’’ pole of the model,
the left \added{(right)} side of the model is coming towards us and identified
with south \added{(north)} for comparison to the data, \added{depending on whether
we follow \citet{2013ApJ...765...85K} or \citet{2008ApJ...687..389S}. Given this uncertainty, we focus on simulating spectra directly toward the continuum source. }

\begin{figure}
\center
\includegraphics[scale=0.3, angle=0]{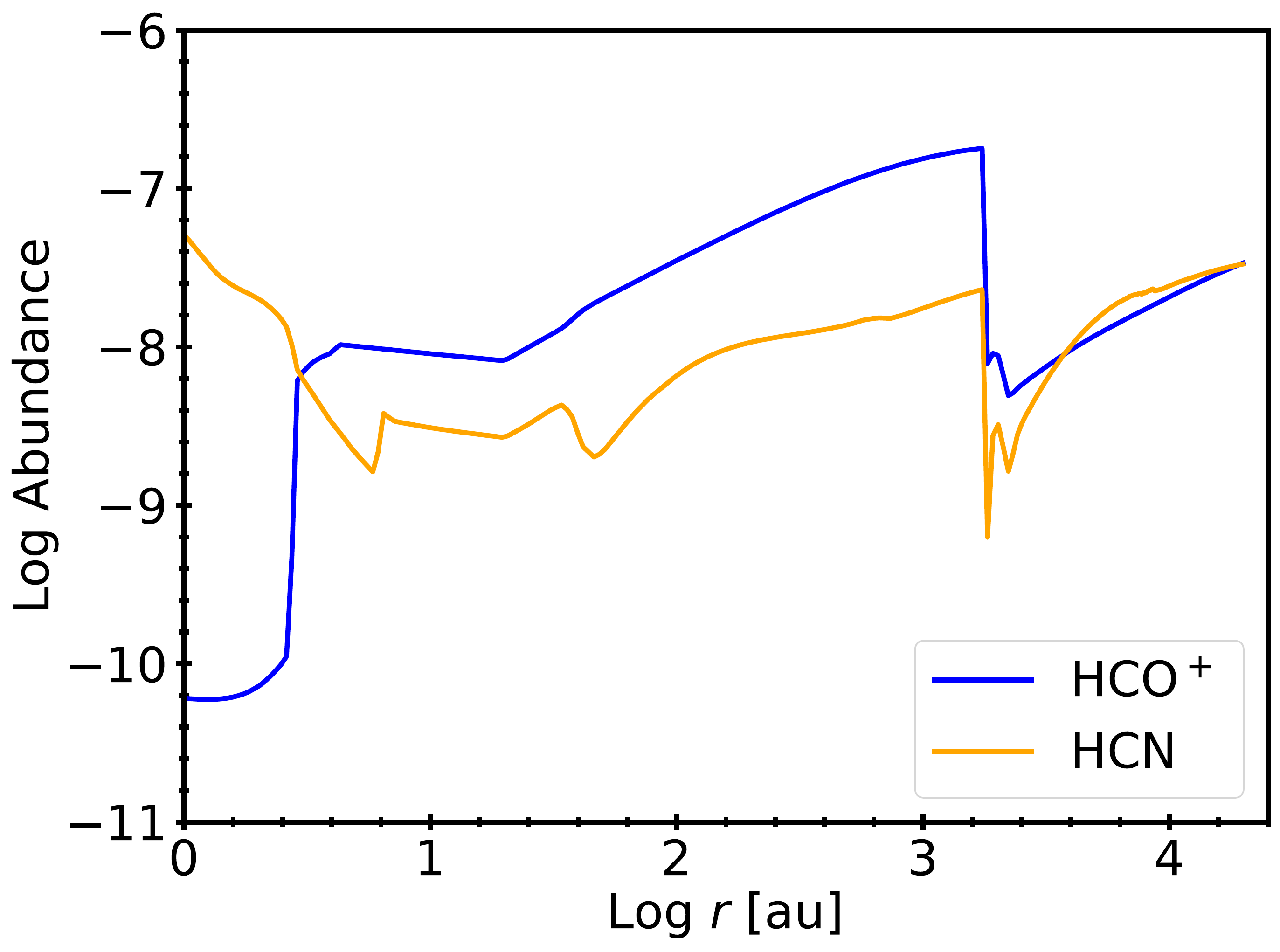}
\caption{
The abundance profile as a function of radius at
$\tcol = 3\ee4$ yr from the chemical model for both \hcop\ and HCN.
}
\label{abun1}
\end{figure}

Chemical models were used to calculate the abundances of species
in the envelope at various ages. The chemical model is described
in detail in  
\citet{2018MNRAS.478.2723Z},
so we provide here only a brief description.
The network includes 21 neutral species, 31 ionic species, and 500 reactions.
Freeze-out and desorption processes are included, but molecular reactions
on the grain surface are not. Sulfur, thus CS, is not included in the model.
\added{We also include all possible charge transfer reactions involving grains. The cosmic-ray ionization rate is set to \eten{-16} s$^{-1}$.}
The abundances were calculated for the tenperature and
density structure of the envelope at a given time. 
\replaced{rather than following the chemistry through the evolution.}{Rather than following the chemistry through the evolution, the abundances were pre-computed for a range of temperatures and densities at a given chemical time to form a look-up table.} The abundances are
produced for a range of radii and polar angles.
Because the abundances were nearly identical
as a function of polar angle, we used the value for 45\degr\ for all
angles and set the abundance to zero in the outflow cavity. Calculations
were available for ages of 1, 3, 4, and 5 times \eten4 yr. 
The abundances used in the models are shown in Figure \ref{abun1}.
These were interpolated onto the grid used by LIME.
These models were computed before the luminosity variation was discovered,
so they assumed the pre-outburst luminosity.

We focus first on the \hcop\ emission, using the
\alma\ Cycle 1 data
\citep{2015ApJ...814...22E},
and the \alma\ Cycle 3 2018 data.
To simulate the continuum source,
we use the LIME-AID code
\citep{2020ApJ...891...61Y}
and insert a continuum source at the center matching the continuum
before deconvolution, as noted above. The position angle was rotated by
90\degree\ because the model has the outflow in the north-south direction.
Then we convolve the
result from LIME-AID to the clean beam used for the data and
extract a spectrum in the same way that we used for the data.
We focus on ages of 1\ee4 to 5\ee4 yr, consistent with the
results from the continuum modeling and past experience with 
the lines.

\begin{figure}
\center
\includegraphics[scale=0.4, angle=0]{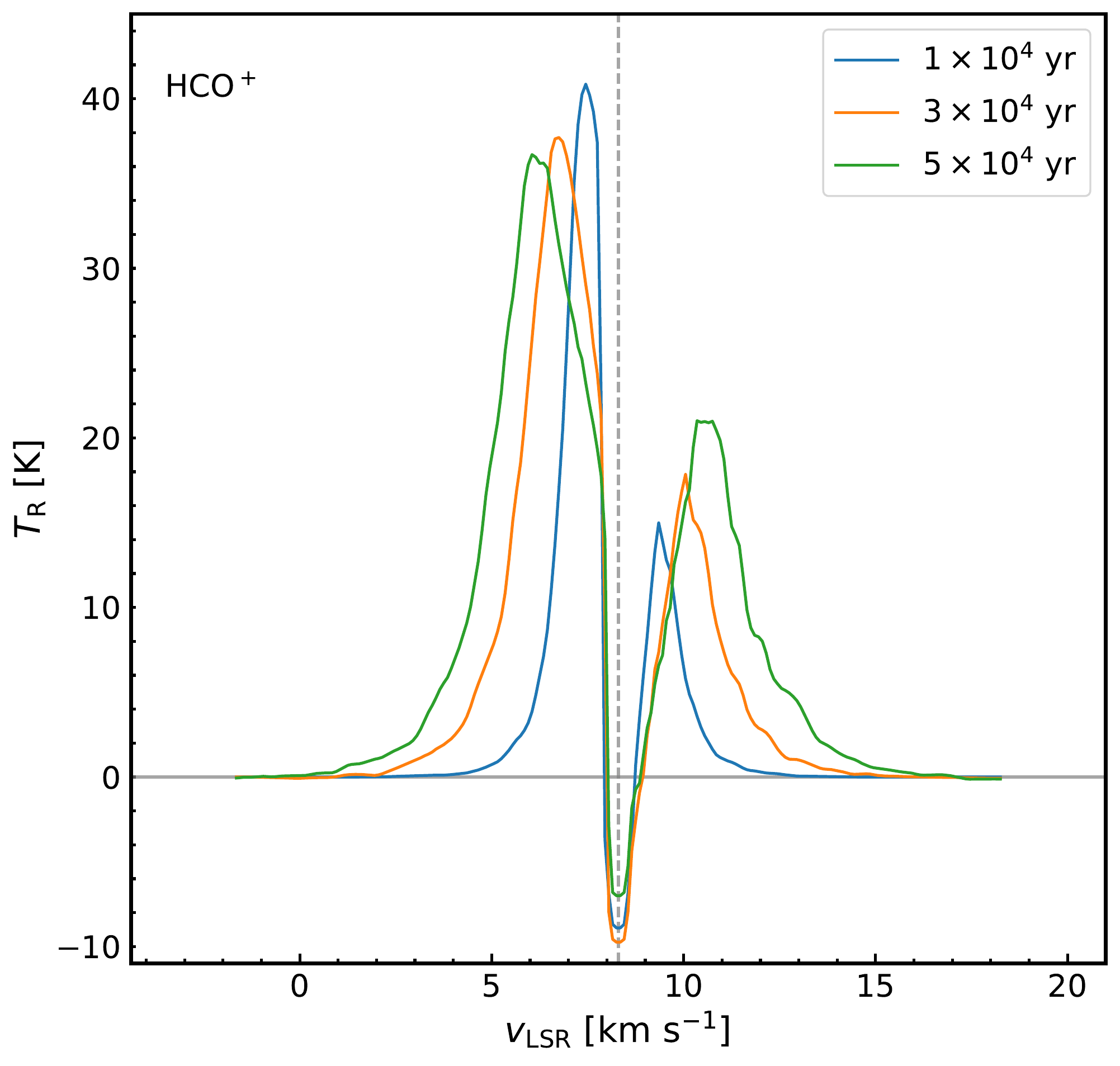}
\includegraphics[scale=0.4, angle=0]{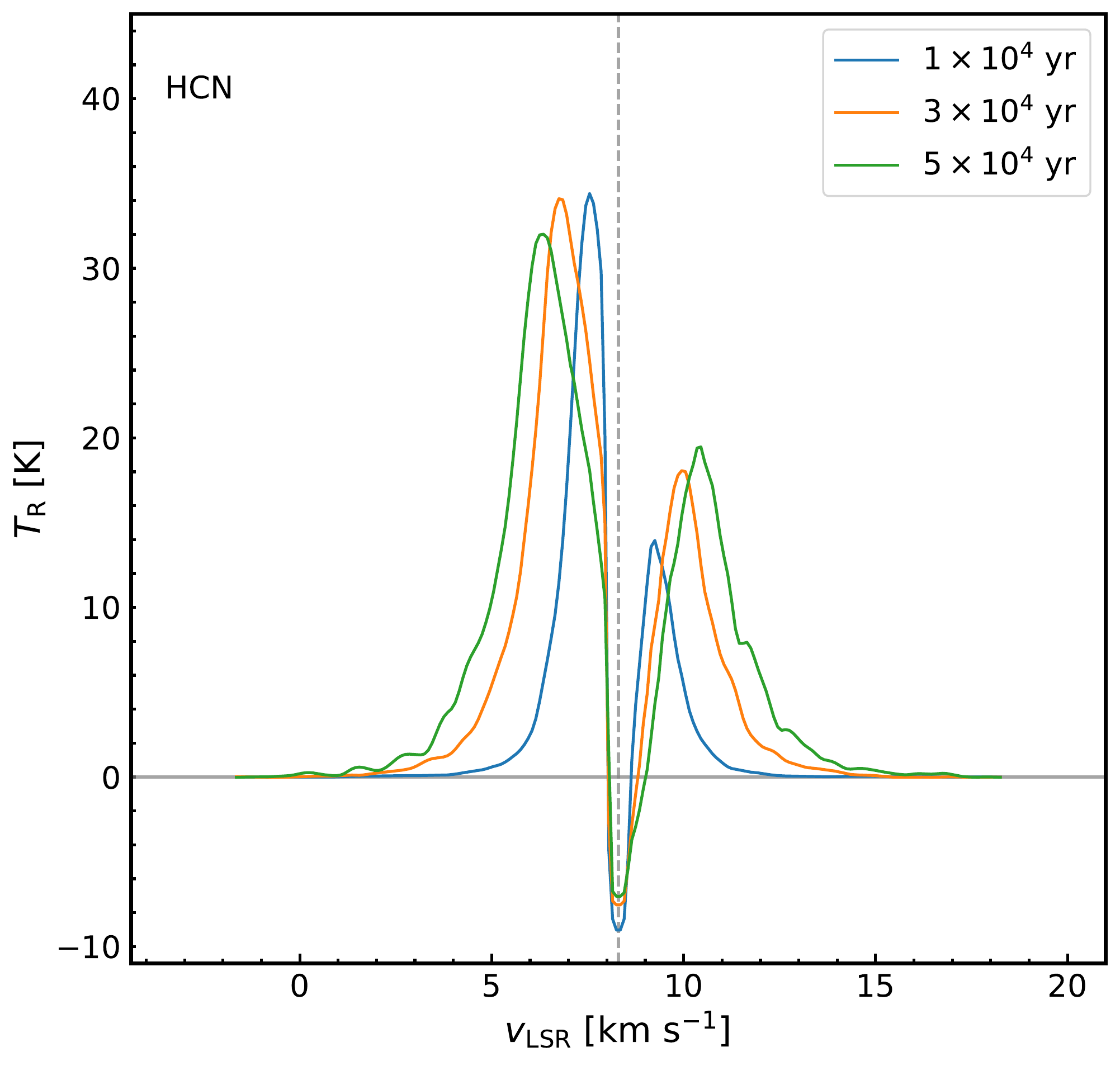}
\caption{
The model  \jj43\ line profiles for ages from $\tcol = 1\ee4$ yr 
to $\tcol = 5\ee4$ yr, using the abundances from the chemical models at those 
ages. 
}
\label{3age-hcop}
\end{figure}

While many things affect the model line profiles, the clearest effect
comes from the age of the system. Figure \ref{3age-hcop} shows line profiles for
\hcop\ and HCN \jj43\ for ages from $\tcol = 1\ee4$ yr to $5\ee4$ yr. As
the collapse proceeds, the infall velocities increase, moving the blue
and red peaks apart, the lines broaden, and the self-absorption in the 
front part of the cloud broadens, eroding the lower velocity emission in 
the red peak.

\begin{figure}
\center
\includegraphics[scale=0.4, angle=0]{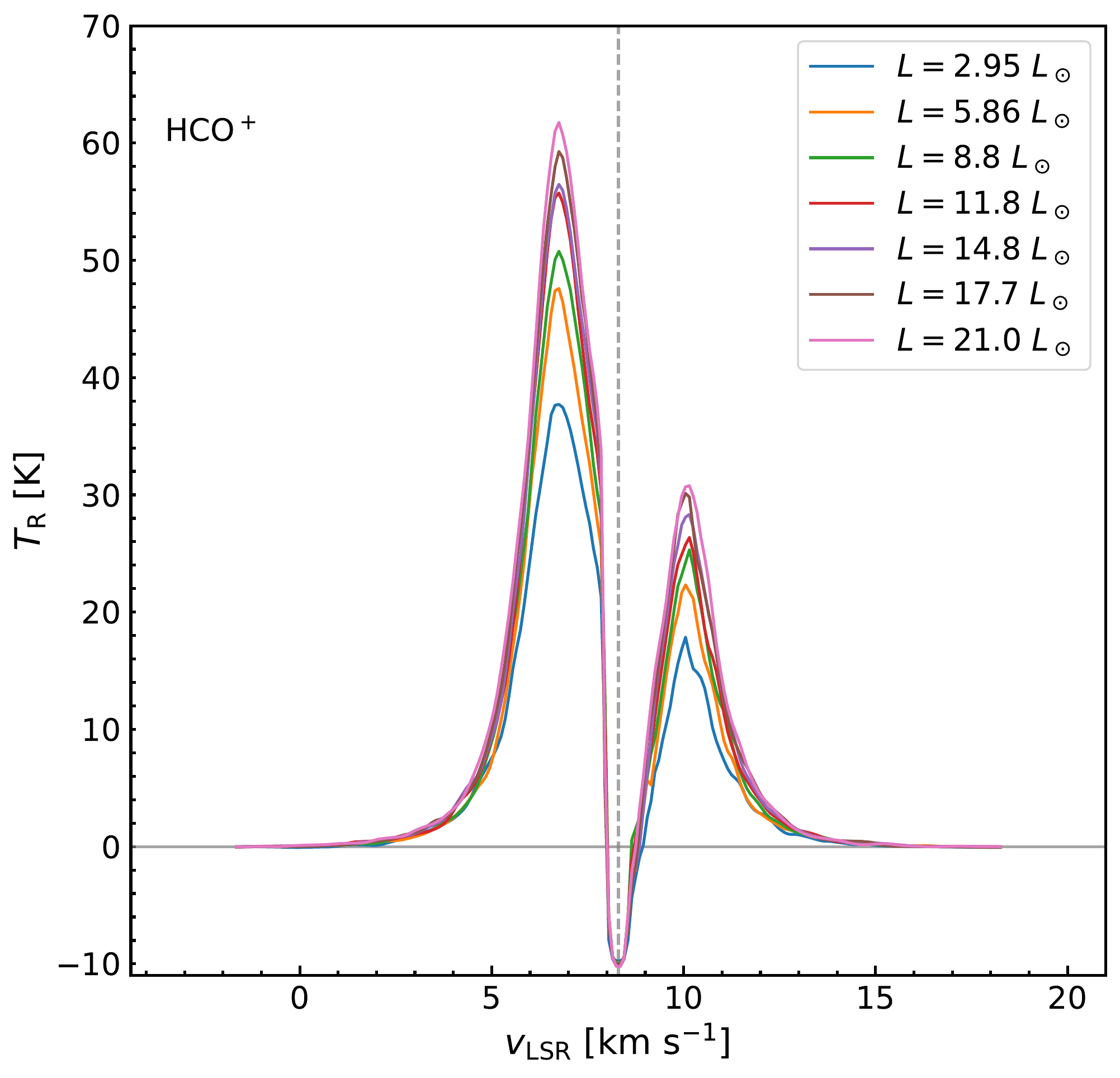}
\caption{
The model \hcop\  \jj43\ line profiles for luminosities from 3 \lsun\ to 21 \lsun, all for $\tcol = 3\ee4$ yr using the abundances from the chemical models at that 
age. 
}
\label{lumdep}
\end{figure}

Because of the luminosity variation, we produced models of the line emission
for a range of luminosities at fixed age ($\tcol = 3\ee4$ yr). These show that the shape of the \hcop\ line is little affected by the source luminosity, but the strength increases smoothly with luminosity (Figure \ref{lumdep}). 
Because the chemical models do not include the luminosity variation, these
line emission models are not self-consistent.

To compare models to observations quantitatively, the figure of merit is 
the absolute value of the difference between observations and models, averaged
over $\pm 3$ \kms\ about the \vlsr.

\begin{equation}
\mean{\mid\rm{Res}\mid} = \mean{\mid \rm{Obs(v)-Model(v)} \mid}
\end{equation}

\begin{figure*}
\center
\includegraphics[scale=0.4, angle=0]{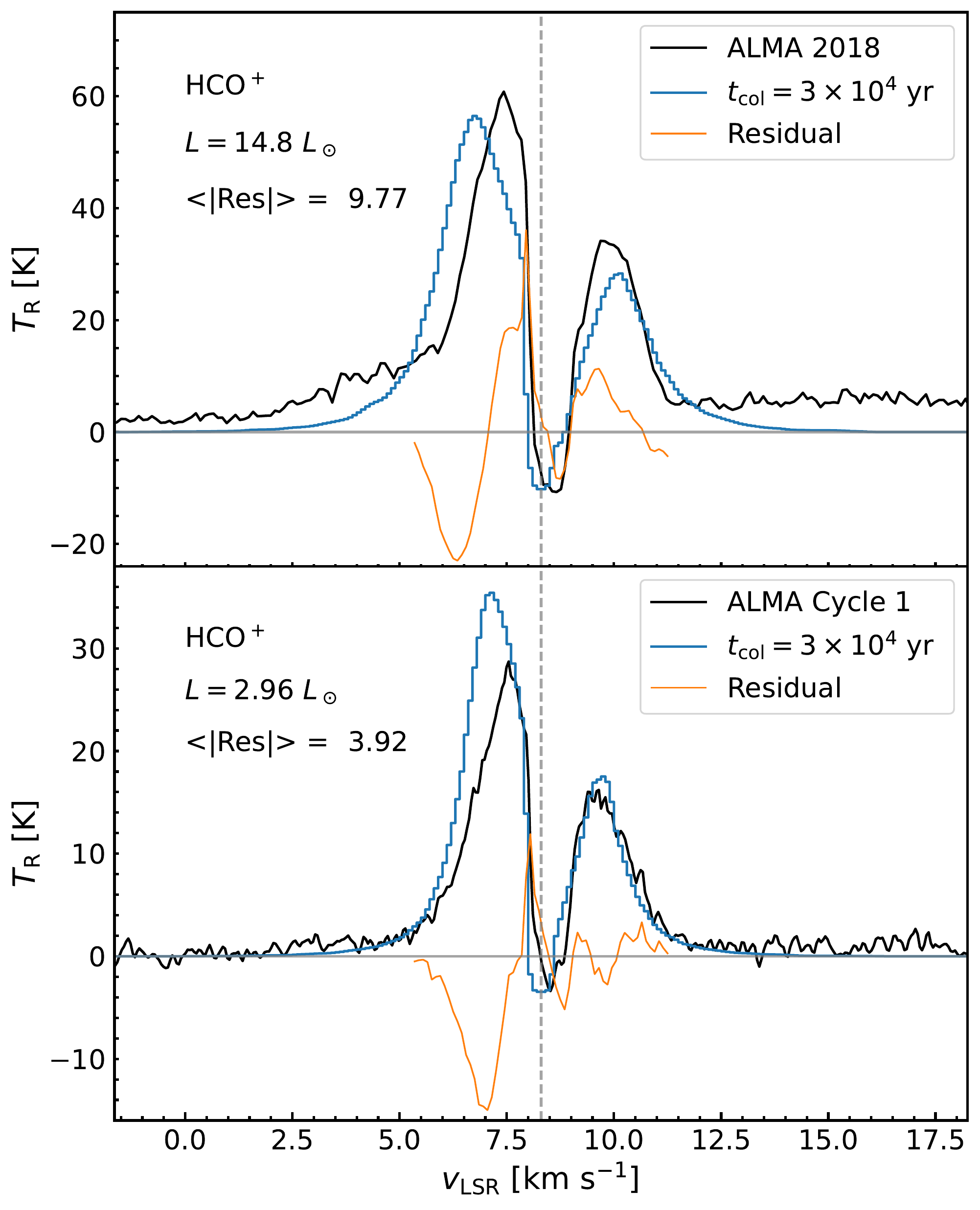}
\includegraphics[scale=0.4, angle=0]{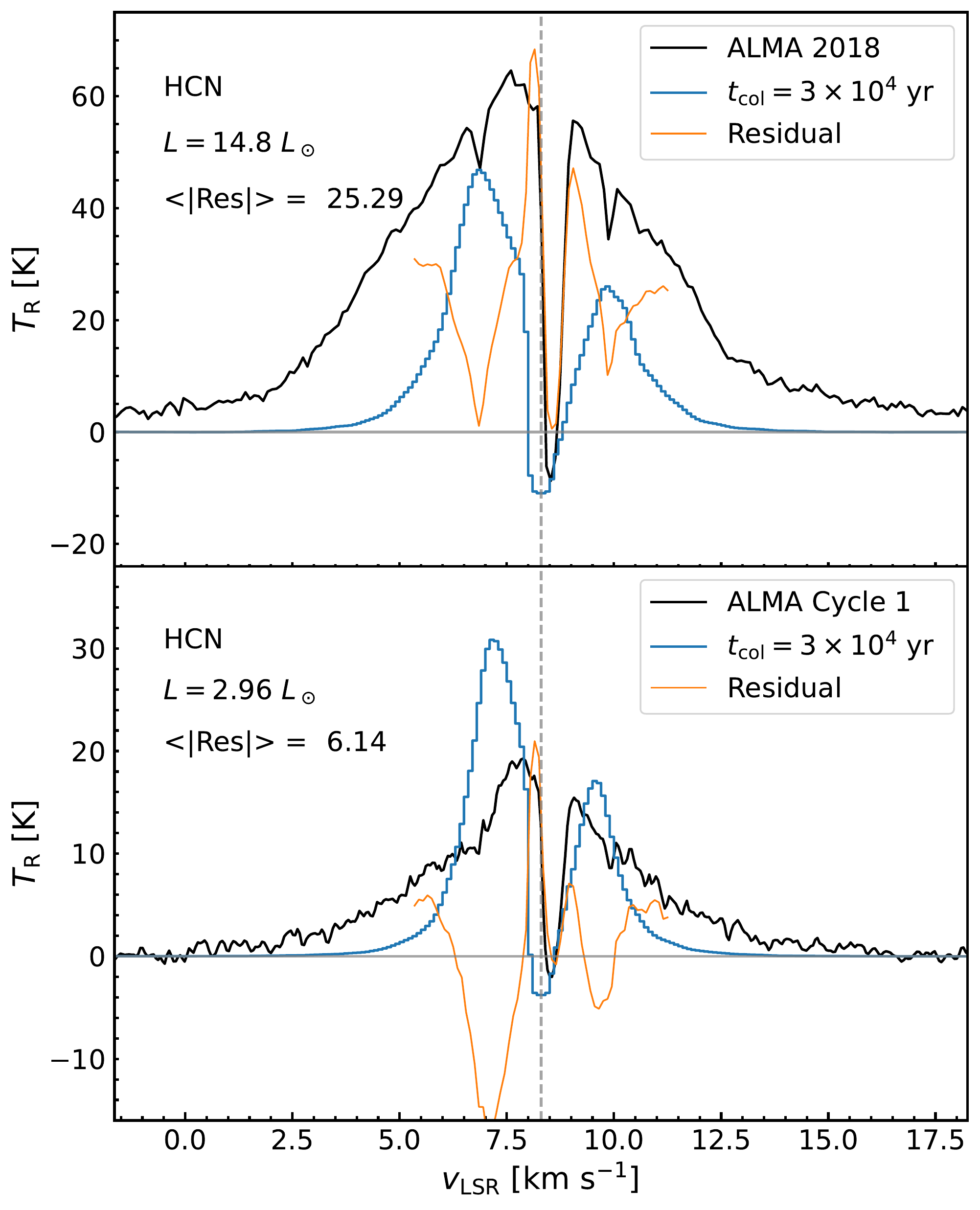}
\caption{
The \hcop\ \jj43\ (left) and HCN (right) line profiles in black from Cycle 3 2018 (top) and Cycle 1 (bottom) with the model for $\tcol = 3\ee4$ yr and the chemical model  in blue. The model for the 2018 data assumes $L = 14.8$ \lsun, while that for Cycle 1 assumes $L = 2.95$ \lsun. The difference between observations and model is plotted in orange and the average absolute value of the difference over velocities from $\vlsr \pm 3$ \kms\ is indicated in the legend.
}
\label{3Dlinechem}
\end{figure*}

The chemical models have only the single variable of age. 
Ages of $1-5\ee4$ yr were tried for both \hcop\ and 
HCN. The age of $\tcol = 3\ee4$ yr was clearly best for \hcop, but
models with $\tcol = 4\ee4$ yr were not much worse.
HCN was about equally (poorly) fitted by 3\ee4 yr and 4\ee4 yr. The resulting
line profiles are shown in Figure \ref{3Dlinechem}. The model can reproduce
the \hcop\ Cycle 1 data reasonably. 
The \added{centroid of the} absorption dip is shifted to the red, matching the observations, but has more absorption near the systemic velocity than the observations. These failings apply to HCN  as well, but
in addition, the modeled HCN lines lack the broad wings seen in the observations. We are not modeling the outflow, a likely source of the broad wings in the observed HCN line.

For the 2018 data, we clearly need to use a larger source luminosity.
Our model based on the W2 photometry indicates $L \approx 20$ \lsun\ when the 2018 data were taken, but this is uncertain as discussed above. We ran a series of models of increasing luminosity (controlled by increasing the stellar radius), while keeping the other variables fixed. The resulting \hcop\ spectra in Figure \ref{lumdep} show that the line increases in strength as the luminosity increases. The model with $L = 15$ \lsun\ matches the 2018 \hcop\ spectrum best. This luminosity worked best if $\tcol = 4\ee4$ yr as well.

\section{Discussion}\label{discussion}

\subsection{Physical Model}\label{bestphysical}

The  3D model that fits the dust continuum emission best
has the properties in Table \ref{best_fit}.
Both the SED and the radial profiles are reasonably well reproduced by this
physical model with $\tcol = 4\ee4$ yr, though ages of $3-5 \times \eten4$ yr are
nearly as good. This result is a major improvement over 1D models, which could not
match the whole SED and required unrealistically young ages to match the
radial profiles. The same basic model, with $\tcol = 3\ee4$ yr, 
supplied with abundances of
\hcop\ from chemical models, gets the basics of the \hcop\ observations correct,
in particular the separation of the blue and red peaks, which is the main
indicator of age. Moving to a 3D model mostly resolved the discrepancy between
models of the continuum and models of the lines. 
The models are, however, not as effective in matching the Cycle 3 data nor the
HCN line profile. 
Overall, models with ages $\tcol = 3-4 \times \eten4$ yr provide the best fits to 
both continuum and line data.

\subsection{Properties of B335}\label{sourceprop}

With the revised distance and best fitting age for the source, we need to
update the basic source properties from those given in
\citet{2015ApJ...814...22E}.
The best fitting ages ($\tcol = 4\ee4$ yr from continuum modeling or
$\tcol = 3\ee4$ yr from line modeling) are generally 
consistent with the outflow age of 2\ee4 yr.
We address first the properties of the source before the luminosity increase.
The new distance results in an {\it observed} luminosity of 1.36 \lsun;
for the adopted inclination angle of 87\degree, the actual central luminosity
must be 2.95 \lsun.
The greater distance also required a higher effective sound speed, $\cseff = 0.30$
\kms, leading to a higher mass inside $\rout = 2\ee4$ au of 3.37 \msun. 
Because the mass infall rate is
proportional to $\cseff^3$, it increases substantially to 6.26\ee{-6} \msunyr.
With an age of $\tcol = 4\ee4$ yr (from the continuum model), the total mass
that has fallen in would be 0.26 \msun. For $\tcol = 3\ee4$ yr (from the line
models), it would be
0.19 \msun. The latter is more consistent with 0.17 \msun\ derived by 
\citet{2019A&A...626A..84E}
(scaled to the new distance) from analysis of the first moment of
various observations (the central blue spot analysis),
though either is probably consistent, given uncertainties.

For the larger mass of 0.26 \msun, the luminosity predicted for the infall
rate is $L = 43.6 f_{acc} $ \lsun. If the radiative efficiency of
accretion $f_{acc} = 0.5$, the predicted luminosity is 21.8 \lsun, 
close to the peak
inferred from modeling the W2 variation. In an episodic accretion picture,
the pre-outburst luminosity resulted from an accumulation of mass in an 
intermediate reservoir, such as a pseudodisk or a Keplerian disk. The luminosity
outburst would then represent a ``catch-up" to match the rate at which envelope
material is infalling.

\subsection{Caveats and Future Prospects}\label{caveats}

While modeling in three dimensions has resolved some puzzles, newer ones are
now apparent. The failure of models to reproduce the continuum emission from
ALMA suggests the existence of additional components, such as a pseudo-disk or a radio jet with
a large positive spectral index. Observations in the 1-10 mm wavelength region 
would resolve the latter possibility. The constraints based on photometry from
3 to 24 \micron\ should be viewed skeptically in light of deep ice absorptions and strong line emission that are emerging from early observations with JWST
\citep{2022arXiv220810673Y}. Contamination of the NEOWISE filters by likely
CO ro-vibrational emission, \hh\ emission lines, atomic lines and ice 
absorptions render suspect the modeling of the luminosity variation in the one data set with \replaced{timely}{time-domain} observations. Chemical models with varying luminosities
combined with the full array of ALMA observations may be able to reconstruct
the luminosity history, and JWST spectra will dramatically improve our 
knowledge of the near-source physical and chemical environment. Future models will need to consider departures from axisymmetry given evidence for infalling streamers.

\section{Summary}

The combination of data from \spitzer, \herschel, and \alma, supplemented by data
from the literature, provides an extremely detailed set of constraints for models
of B335. Models of the continuum emission with rotating, infalling (TSC) envelopes, outflow cavities, and disks can provide a reasonable match to the SED and radial profiles at submillimeter wavelengths for ages (\tcol) of $3-4 \ee4$ yr. Similar ages with chemical models provide the best fit to the \hcop\ spectra from \alma. The HCN spectrum is, however, less well-fitted.

B335 has undergone an increase in luminosity over the last few years of a factor of 5-7, but is now decreasing back toward its previous luminosity of about 3 \lsun. The \alma\ observations at various times during this luminosity excursion show strong responses in the line strength with increasing luminosity. There were also pronounced increases in emission from COMs, which will be analyzed in a separate paper.

The revised infall rate predicts a luminosity near the peak of the recent outburst, suggesting that the pre-outburst source was storing matter in a structure between the envelope and the star.

\begin{acknowledgments}
This paper makes use of the following \alma\ data: ADS/JAO.ALMA\#2012.1.00346.S and 2015.1.00169.S. 
\alma\ is a partnership of ESO (representing its member states), NSF (USA) and 
NINS (Japan), together with NRC (Canada), MOST and ASIAA (Taiwan), and KASI (Republic of Korea), in 
cooperation with the Republic of Chile. The Joint \alma\ Observatory is 
operated by ESO, AUI/NRAO and NAOJ. 
The National Radio Astronomy Observatory is a facility of the National Science Foundation operated under cooperative agreement by Associated Universities, Inc. 
This work has made use of data from the European Space Agency (ESA) mission
{\it Gaia} (\url{https://www.cosmos.esa.int/gaia}), processed by the {\it Gaia}
Data Processing and Analysis Consortium (DPAC,
\url{https://www.cosmos.esa.int/web/gaia/dpac/consortium}). Funding for the DPAC
has been provided by national institutions, in particular the institutions
participating in the {\it Gaia} Multilateral Agreement.
All spectral line data were taken from the Spectral Line 
Atlas of Interstellar Molecules (SLAIM) 
(Available at http://www.splatalogue.net). 
NJE thanks the Astronomy Department of the University of Texas for
research support.
The research of JKJ is supported by a grant from the Independent Research Fund Denmark (grant No. 0135-00123B). J.-E. Lee was supported by the Basic Science Research 
Program through the National Research Foundation of Korea (NRF) 
funded by the Ministry of Education of the Korean government 
(grant No. NRF-2012R1A1A2044689).
\end{acknowledgments}

\software{astropy
\citep{2013A&A...558A..33A,2018AJ....156..123A, 2022arXiv220614220T},
GILDAS
\citep{2005sf2a.conf..721P,2013ascl.soft05010G}
}



\bibliographystyle{aasjournal}
\clearpage
\bibliography{more,jdf.infall}


\end{document}